\documentclass[twocolumn,astrosymb]{aastex631}
\usepackage[flushleft]{threeparttable}
\usepackage{supertabular}

\newcommand{\pipe}{\texttt{pipeline}}
\begin{document}

\title{COSMOS-Web: MIRI Data Reduction and Number Counts at 7.7$\mu$m  using JWST}

\author[0000-0003-0129-2079]{Santosh Harish}
\affiliation{Laboratory for Multiwavelength Astrophysics, School of Physics and Astronomy, Rochester Institute of Technology, 84 Lomb Memorial Drive, Rochester, NY 14623, USA}

\author[0000-0001-9187-3605]{Jeyhan S. Kartaltepe}
\affiliation{Laboratory for Multiwavelength Astrophysics, School of Physics and Astronomy, Rochester Institute of Technology, 84 Lomb Memorial Drive, Rochester, NY 14623, USA}

\author[0000-0001-9773-7479]{Daizhong Liu}
\affiliation{Purple Mountain Observatory, Chinese Academy of Sciences, 10 Yuanhua Road, Nanjing 210023, China}

\author[0000-0002-6610-2048]{Anton M. Koekemoer}
\affiliation{Space Telescope Science Institute, 3700 San Martin Dr., Baltimore, MD 21218, USA}

\author[0000-0002-0930-6466]{Caitlin M. Casey}
\affiliation{Department of Physics, University of California, Santa Barbara, Santa Barbara, CA 93109, USA}
\affiliation{The University of Texas at Austin, 2515 Speedway Blvd Stop C1400, Austin, TX 78712, USA}
\affiliation{Cosmic Dawn Center (DAWN), Denmark}

\author[0000-0002-3560-8599]{Maximilien Franco}
\affiliation{Université Paris-Saclay, Université Paris Cité, CEA, CNRS, AIM, 91191 Gif-sur-Yvette, France}
\affiliation{The University of Texas at Austin, 2515 Speedway Blvd Stop C1400, Austin, TX 78712, USA}

\author[0000-0003-3596-8794]{Hollis B. Akins}
\affiliation{The University of Texas at Austin, 2515 Speedway Blvd Stop C1400, Austin, TX 78712, USA}

\author[0000-0002-7303-4397]{Olivier Ilbert}
\affiliation{Aix Marseille Univ, CNRS, CNES, LAM, Marseille, France }

\author[0000-0002-7087-0701]{Marko Shuntov}
\affiliation{Cosmic Dawn Center (DAWN), Denmark} 
\affiliation{Niels Bohr Institute, University of Copenhagen, Jagtvej 128, DK-2200, Copenhagen, Denmark}

\author[0000-0003-4761-2197]{Nicole E. Drakos}
\affiliation{Department of Physics and Astronomy, University of Hawaii, Hilo, 200 W Kawili St, Hilo, HI 96720, USA}

\author[0000-0003-0209-674X]{Mike Engesser}
\affiliation{Space Telescope Science Institute, Baltimore, MD 21218, USA}

\author[0000-0002-9382-9832]{Andreas L. Faisst}
\affiliation{Caltech/IPAC, MS 314-6, 1200 E. California Blvd. Pasadena, CA 91125, USA}

\author[0000-0002-0236-919X]{Ghassem Gozaliasl}
\affiliation{Department of Computer Science, Aalto University, P.O. Box 15400, FI-00076 Espoo, Finland}
\affiliation{Department of Physics, University of, P.O. Box 64, FI-00014 Helsinki, Finland}

\author[0000-0001-9189-7818]{Crystal L. Martin}
\affil{Department of Physics, University of California, Santa Barbara, Santa Barbara, CA 93109, USA}

\author[0000-0002-3301-3321]{Michaela Hirschmann}
\affiliation{Institute of Physics, GalSpec, Ecole Polytechnique Federale de Lausanne, Observatoire de Sauverny, Chemin Pegasi 51, 1290 Versoix, Switzerland}
\affiliation{INAF, Astronomical Observatory of Trieste, Via Tiepolo 11, 34131 Trieste, Italy}

\author[0000-0002-5588-9156]{Vasily Kokorev}
\affiliation{Kapteyn Astronomical Institute, University of Groningen, PO Box 800, 9700 AV Groningen, The Netherlands}

\author[0000-0003-3216-7190]{Erini Lambrides}\altaffiliation{NPP Fellow}
\affiliation{NASA-Goddard Space Flight Center, Code 662, Greenbelt, MD, 20771, USA}

\author[0000-0002-9489-7765]{Henry Joy McCracken}
\affiliation{Institut d’Astrophysique de Paris, UMR 7095, CNRS, and Sorbonne Université, 98 bis boulevard Arago, F-75014 Paris, France}

\author[0000-0002-6149-8178]{Jed McKinney}
\altaffiliation{NASA Hubble Fellow}
\affiliation{Department of Astronomy, The University of Texas at Austin, 2515
Speedway Blvd Stop C1400, Austin, TX 78712, USA}

\author[0000-0003-2397-0360]{Louise Paquereau} 
\affiliation{Institut d’Astrophysique de Paris, UMR 7095, CNRS, and Sorbonne Université, 98 bis boulevard Arago, F-75014 Paris, France}

\author[0000-0002-4485-8549]{Jason Rhodes}
\affiliation{Jet Propulsion Laboratory, California Institute of Technology, 4800 Oak Grove Drive, Pasadena, CA 91001, USA}

\author[0000-0002-4271-0364]{Brant E. Robertson}
\affiliation{Department of Astronomy and Astrophysics, University of California, Santa Cruz, 1156 High Street, Santa Cruz, CA 95064, USA}



\begin{abstract}
The COSMOS-Web survey is the largest JWST Cycle 1 General Observer program covering a contiguous $\sim$0.54 deg$^2$ area with NIRCam imaging in four broad-band filters and a non-contiguous $\sim$ 0.2 deg$^2$ with parallel MIRI imaging in a single broad-band filter, F770W. Here we present a comprehensive overview of the MIRI imaging observations, the data reduction procedure, the COSMOS-Web MIRI photometric catalog, and the first data release including the entire COSMOS-Web MIRI coverage. Data reduction is predominantly based on the JWST Science Calibration Pipeline with an additional step involving custom background subtraction to mitigate the presence of strong instrumental features and sky background in the MIRI images. We reach 5$\sigma$ (point source) limiting depths ($m_\textnormal{F770W} \sim 25.51$ based on $r\sim0.3\arcsec$ circular apertures) that are significantly better than initial expectations. We create a COSMOS-Web MIRI catalog based on the images presented in this release and compare the F770W flux densities with the \emph{Spitzer}/IRAC CH4 measurements from the COSMOS2020 catalog for CH4 detections with S/N $>5$. We find that these are in reasonable agreement with a small median offset of $<0.05$ mag. We also derive robust 7.7$\mu$m number counts spanning five orders of magnitude in flux ($\sim$0.2-2300 $\mu$Jy) -- making COSMOS-Web the only JWST survey to date to efficiently sample such a large flux range -- which is in good agreement with estimates from other JWST and IRAC surveys.
\end{abstract}



\section{Introduction} \label{sec:intro}

Mid-infrared (MIR) imaging plays a critical role in extragalactic astronomy by providing a unique perspective on distant galaxies and their intricate structures. In the context of star formation, much of the UV/optical emission is absorbed by dust in the interstellar medium (ISM) and re-radiated at longer wavelengths, particularly in the IR regime (8-1000\,$\mu$m). Additionally, even for galaxies with modest dust extinction, the strength of polycyclic aromatic hydrocarbons (PAHs; \citealt{Leger1984,Genzel1998}) emission (3-12\,$\mu$m) is considered a good indicator of obscured star formation (e.g., \citealt{Wu2005}) for galaxies up to $z\sim2$ \citep{Houck2005,Papovich2006}. 

IR measurements are essential for deriving accurate stellar masses at higher redshifts, where they probe the rest-frame optical to near-infrared (NIR) emission of galaxies, tracing the older stellar population. The MIR wavelength regime is also crucial for identifying obscured active galactic nuclei (AGN) because the hot dust torus emission of an AGN peaks in this spectral regime, which is distinct from the host galaxy emission at these wavelengths (see review by \citealt{Hickox2018}). For galaxies with high dust and gas content, MIR-based selection techniques (e.g., \citealt{Kirkpatrick2012}) are efficient at identifying hidden AGN population that are missed by other wavelengths, especially those blueward of the UV/optical wavelengths.

The lack of sensitivity of ground-based telescopes to MIR emission coupled with strong absorption in the terrestrial atmosphere means that space-based telescopes are the most efficient means of mapping the universe in the MIR. Several IR missions including the Infrared Astronomical Satellite (IRAS; \citealt{Neugebauer1984}), Infrared Space Observatory (ISO; \citealt{Kessler1996}), Spitzer Space Telescope (Spitzer; \citealt{Werner2004}), AKARI \citep{Murakami2007}, and Wide-field Infrared Survey Explorer (WISE; \citealt{Wright2010}) have successfully explored a variety of dusty star forming galaxy populations including the luminous infrared galaxies (LIRGs; $L_{IR}>10^{11}L\odot$) and ultra-luminous infrared galaxies (ULIRGs; $L_{IR}>10^{12}L\odot$), which are rare in the local Universe but increasingly important at $z>1$ \citep{Sanders1996,LeFloch2005}. Notable among these is the Spitzer mission, which has performed several deep-field surveys to probe the faint IR populations, especially using the Infrared Array Camera (IRAC) instrument \citep{Fazio2004a}. IRAC has played a crucial role in confirming some of the highest redshift objects identified using the Hubble Space Telescope (HST), including the well-studied object GN-z11 \citep{Oesch2016}, and constraining their stellar masses, ages, and star formation histories (e.g., \citealt{Stark2007,Labbe2010,Labbe2013}). 

Building on the rich legacy of previous space-based IR missions, the successful launch of the James Webb Space Telescope (JWST; \citealt{Gardner2006}) has enabled unprecedented observations in the NIR and MIR, with sensitivity and resolution far superior to its predecessors. In particular, the Mid-Infrared Instrument (MIRI; \citealt{Rieke2015,Wright2023}) represents a significant advancement in its ability to probe the 5-28$\mu$m range through imaging, spectroscopy, and coronagraphy. Because longer wavelengths are less affected by dust extinction, MIRI offers several key advantages, including: (1) characterization of stellar populations and star formation histories (SFHs) for galaxies at $z>5$ (e.g., \citealt{Papovich2023}), as the ability to detect evolved stellar populations improves at redder wavelengths, (2) detection of H$\alpha$ emission line galaxies at $z>6.6$ (e.g., \citealt{Rinaldi2023}), which is a valuable probe of star formation and AGN in the epoch of reionization (EoR), (3) detection of massive starbursts at $z>4$ (e.g., \citealt{Alvarez2023}), as MIRI can probe the rest-NIR ($\lambda>1\mu$m), (4) detection of AGN in massive galaxies at $z>1$ (e.g., \citealt{Lyu2024}), with MIRI uniquely able to observe the hot dusty torus emission that dominates the spectral energy distribution (SED) in the rest-NIR ($\lambda > 1$ µm).

Surpassing early expectations, MIRI has already demonstrated its prowess by confirming some of the most distant galaxies discovered by JWST, including JADES-GS-z14-0 at $z=14.32$ \citep{Helton2025} and GHZ2/GLASS-z12 at $z=12.33$ \citep{Zavala2025}, with the highest redshift confirmation to date when the universe was just 290 million years old \citep{Carniani2024}.  In addition to these high-redshift discoveries, MIRI has been instrumental in detecting a large population of obscured AGN \citep{Yang2023a} up to $z \sim 8$ \citep{Lyu2024}, identifying exotic objects such as NIR-dark, MIR-only detections \citep{Perez-Gonzalez2024b}, and enabling accurate stellar mass estimates for the so-called ``little red dots" \citep{Akins2023, Perez-Gonzalez2024a, Akins2024} by providing critical information in the MIR.

The COSMOS-Web (PID 1727; PIs: Kartaltepe \& Casey; \citealt{Casey2023}) survey is the largest JWST Cycle 1 GO program, both in terms of area on the sky and total prime time allocation, mapping the central region of the COSMOS field \citep{Koekemoer2007,Scoville2007,Capak2007} covering a large contiguous area of 0.54 deg$^2$ using four Near-Infrared Camera (NIRCam; \citealt{Rieke2005,Rieke2023}) filters (F115W, F150W, F277W, and F444W) and a non-contiguous area of 0.2 deg$^2$ using a single MIRI filter (F770W). An in-depth overview of the survey design, primary science goals, and comparison with other deep field programs from JWST's first year of observations is presented by \cite{Casey2023}.

This paper describes the MIRI imaging component of COSMOS-Web, including details of the completed observations in Section \ref{sec:obs}, an overview of the data reduction using the JWST science calibration pipeline and the survey depth in Section \ref{sec:redux}. The construction of a COSMOS-Web MIRI source catalog based on F770W detections is described in Section \ref{sec:source_cat} together with estimates of source completeness. We present a comparison with the IRAC CH4 photometry, the photometric redshift distribution of the 7.7$\mu$m sources, and the number counts at 7.7$\mu$m including comparisons with other estimates based on JWST and Spitzer observations as well as model predictions in Section \ref{sec:results}, followed by a summary in Section \ref{sec:summary}.

\section{Observations} \label{sec:obs}

The principal design of COSMOS-Web was driven by the NIRCam observations, which consist of a  contiguous 41.5 $' \times$ 46.6$'$ rectangular mosaic (oriented at an angle 20$^{\circ}$ relative to North) centered at RA = 10:00:27.92, Dec = $+$02:12:03.5, using the aforementioned NIRCam filters (discussed in a separate paper; Franco et al., in prep.), and complemented by parallel MIRI imaging using the F770W filter. The F770W filter was chosen to accurately constrain stellar masses at $z>4$ due to its sensitivity and the greater wavelength separation it provides over F444W, offering better differentiation than filters like F560W. A typical MIRI visit consisting of eight exposures was obtained through two separate executions of the 4TIGHT dither pattern at the same mosaic position (for more details, see \citealt{Casey2023}). The MIRI observations consist of 160 distinct visits with each visit containing anywhere between two to eight exposures (263s per exposure), depending on whether the visit was fully or partially observed. 

Observations were taken over four different epochs between January 2023 and May 2024. For the most part, the observing program was executed as planned; however, a small fraction of the total visits ($\sim$10\%) were either fully skipped or partially observed due to guide star acquisition issues, and those visits were re-observed in the final two epochs. This resulted in a slightly larger area covered than initially proposed, with some parts of the mosaic having fewer exposures due to the nature of parallel observations obtained at different position angles (P.A.). 

Considering the field-of-view of both the main and the Lyot coronagraphic imager (which has the same optical path as the main imager\footnote{\url{https://jwst-docs.stsci.edu/jwst-mid-infrared-instrument/miri-observing-modes/miri-imaging\#MIRIImaging-Imagingfieldofview}}), the total MIRI area coverage per visit is $\sim$4.5 arcmin$^2$, and therefore the total area covered with MIRI (accounting for both additional coverage from repeat visits and reduced coverage due to skipped/failed visits) is 722 arcmin$^2$ or $\sim$0.2 deg$^2$. Figure \ref{fig:MIRI_visits_map} shows a map of all the completed MIRI visits from the observing program, and information regarding these visits are provided in Table \ref{tab:visits} in the appendix.

\section{Data Reduction}  \label{sec:redux}

The MIRI imaging data were reduced using the standard JWST Science Calibration Pipeline\footnote{\url{https://github.com/spacetelescope/jwst}} (v1.12.5; \citealt{Bushouse2023}), hereafter referred to as \pipe, employing calibrations based on the JWST Calibration Reference Data System\footnote{\url{https://jwst-crds.stsci.edu/}} Pipeline Context version 1130. The reduction procedure, starting from raw uncalibrated data to the creation of final mosaics through three different stages, including the implementation of a custom routine for improved background subtraction, is detailed in the following subsections.

\subsection{Stage 1: Detector Processing} \label{sec:stage1}
The \texttt{Detector1Pipeline} a.k.a. \textsc{calbwebb\_detector1} performs detector-level processing that reduces raw non-destructively read ramps to uncalibrated slope (\emph{rate}) images. This primarily includes data quality initialization, reference pixel correction, jump detection, slope fitting along with MIRI-specific steps such as corrections for reset anomaly, first and last frame (all due to transient effects), non-linearity, ``reset switch charge decay'' and dark subtraction. For a typical visit, the default stage~1 parameters are sufficient without the need for modifications. However, we note that the reference pixel and reset corrections are not applied by default in recent versions of the \pipe, as these effects are already incorporated into the dark reference file (Mike Regan, priv. comm.). This will be implemented accordingly in a future data release. 

Several artifacts arising due to certain non-ideal behaviors of the detector (see \citealt{Morrison2023} for a detailed discussion) remain unaddressed in the \pipe. A visual inspection of the \emph{rate} images indicates the presence of some of these artifacts, albeit in a smaller number of visits. The most commonly observed artifact is called a `cosmic-ray shower,' which is the result of large clusters of energetic particles produced by cosmic rays, affecting a large number of pixels on the MIRI detectors, typically below the detection threshold for a direct cosmic ray hit. In contrast to the `snowball' artifact observed in the case of NIRCam, the cosmic ray showers are non-circular in nature and lack a well-defined saturated core. Improvements to the \pipe\ to correct these artifacts are currently under development, and at the time of this writing, we note that a first-order correction for cosmic-ray showers is included in the jump detection step of the \pipe, which will be implemented in a future data release.

\subsection{Stage 2: Image Processing} \label{sec:stage2}
Following the initial corrections applied in Stage~1, we further processed the \emph{rate} images using \texttt{Image2Pipeline}, a.k.a \textsc{calbwebb\_image2}, which was executed with the default settings. This crucial stage involves the assignment of World Coordinate System (WCS) information, flat-fielding, and photometric calibration, culminating in fully calibrated individual exposures. Thus, each rate map from Stage 1 is transformed into a calibrated science (\emph{cal}) image, with units converted from counts per second to MJy/sr.

Although most MIRI visits have coverage with four or more exposures, a small but significant fraction ($\sim$13\%) of them have only two exposures. As reported by \citep{Bagley2023} in the CEERS program, the background levels in the final mosaics are significantly improved if the \emph{SkyMatch} step is run on each \emph{cal} image individually before proceeding with the final stage of \pipe. The \emph{SkyMatch} step computes the sky values in a given image and subtracts the measured background (typically the median value) from said image. For this step, we set \texttt{skymethod = `local'} and \texttt{subtract = True} to subtract a constant sky background calculated for each input image. 

\begin{figure*}[htp]
	\centering
	\hspace*{-1em}\includegraphics[width=2.2\columnwidth]{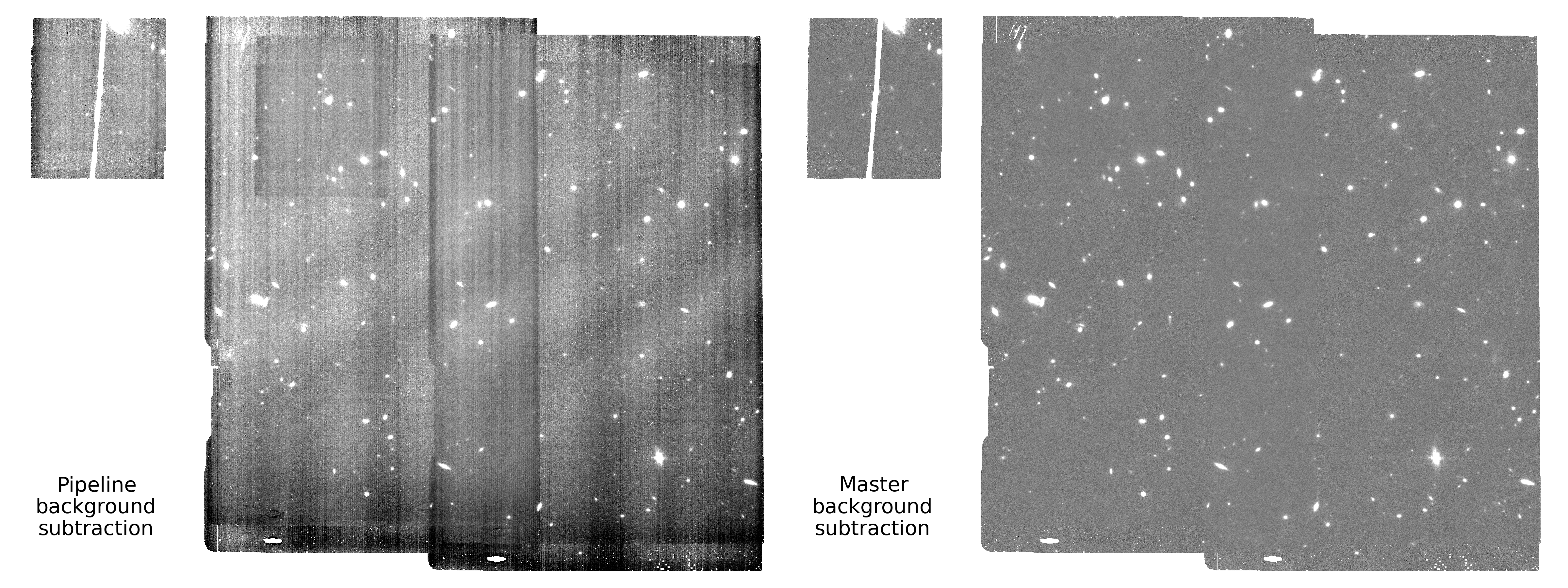}
	\caption{Example MIRI F770W image based on the default pipeline output (\emph{left}) and the same with our \emph{master background subtraction} procedure applied (\emph{right}) as detailed in Section \ref{sec:stage3}. The stretch in both images are the same.}
	\label{fig:bkg_subt}
\end{figure*}

Upon inspecting the background-subtracted \emph{cal} image, we noted significant residuals in the background (see Figure \ref{fig:bkg_subt}), similar to those reported in other JWST MIRI observations \citep[e.g.,][]{Yang2023b,Perez-Gonzalez2024a,Alberts2024}. In the following text, we discuss our custom background subtraction procedure to mitigate this issue, similar in principle to those implemented by these other surveys. 

\subsubsection{Master Background Subtraction}
More than two years into science operations, backgrounds in MIRI are fairly well characterized, which are expected to be a combination of sky background and detector effects. The former includes emission with astronomical origins (mainly zodiacal dust and Milky Way interstellar medium) predominantly for wavelengths $\sim5-13\,\mu$m, while redder wavelengths are dominated by thermal self-emission and stray light from the observatory sunshield \citep{Glasse2015,Rigby2023}. The detector effects include artifacts arising from row-column effects, persistence, the cruciform effect, ``tree-rings'' (for more details see \citealt{Morrison2023,Dicken2024}), to name a few. At the time of this writing, some workarounds are available \footnote{\url{https://jwst-docs.stsci.edu/known-issues-with-jwst-data/miri-known-issues/miri-imaging-known-issues\#MIRIImagingKnownIssues-summarySummaryofCommonIssuesandWorkarounds}} to mitigate some of these issues (as additional steps outside of the main pipeline processing); however, we found that these have minimal impact in terms of removing the large-scale gradients in our image backgrounds. Therefore, we address this issue using a custom routine called \emph{master background subtraction}\footnote{\url{https://github.com/1054/Crab.Toolkit.JWST/blob/main/docs/reprocessing_MIRI_with_ImageForImageBackgroundSubtraction.md}} (similar to the ``super-background'' strategy adopted by other JWST surveys, e.g., \citealt{Yang2023b,Perez-Gonzalez2024a,Alberts2024}), which helps produce final mosaic images with a uniform sky background.

As an initial step, we perform a full reduction using the standard \pipe~producing the final Stage 3 mosaic image (see next subsection) for each individual MIRI visit. Subsequently, we used Source Extractor (\textsc{SE}; \citealt{Bertin1996}, version 2.28.0) to detect bright source emission in this mosaic image and create a source mask called the ``galaxy seed image '' for each MIRI visit. We then apply this source mask to each \emph{rate} image in the corresponding MIRI visit, producing a ``source-masked rate'' image. Fortunately, most of the COSMOS-Web visits were observed in contiguous blocks in time (except for the repeat visits from Apr-May 2024), where each block contained at least five or more visits observed in succession. Using this to our advantage, we merged the ``source-masked rate'' images of all other exposures from the same visit, as well as exposures from other contemporaneous visits (at least two), forming a visit set to produce a unique ``merged source-masked rate'' image (since this will always exclude the exposure at hand but merge all other exposures/visits). This ensures that the large-scale temporal variation of the 2D background is well captured by merging multiple exposures/visits observed contemporaneously. Care was taken to ensure that holes (where source emission was masked) caused due to dithering are replaced with median values while merging. We then re-run Stage 2 on the original \emph{rate} image, where the corresponding ``merged source-masked rate'' image is subtracted using the \texttt{Image2Pipeline.bkg\_subtract} routine to produce a \emph{master-background} subtracted \emph{cal} image. This procedure was repeated set-by-set for every exposure in each MIRI visit in a set.

Finally, the resulting \emph{master-background} subtracted \emph{cal} images are fed to the next stage of pipeline processing (Stage 3). Figure \ref{fig:bkg_subt} shows an example F770W image before and after applying our \emph{master-background subtraction} procedure.

\subsection{Stage 3: Mosaic Creation} \label{sec:stage3}
The final stage of the reduction, \texttt{Image3Pipeline}, a.k.a \textsc{calbwebb\_image3} combines the calibrated data from Stage 2 from multiple exposures (usually a dither pattern) into a single, distortion-corrected science-ready mosaic image. This stage also includes some additional steps performed on the individual exposures, such as astrometric alignment and outlier rejection, before they are resampled and combined to produce a mosaic image. We briefly describe our adopted strategy with respect to these steps in the following text.

\begin{figure}[htp]
	\centering
	\includegraphics[width=1.1\columnwidth]{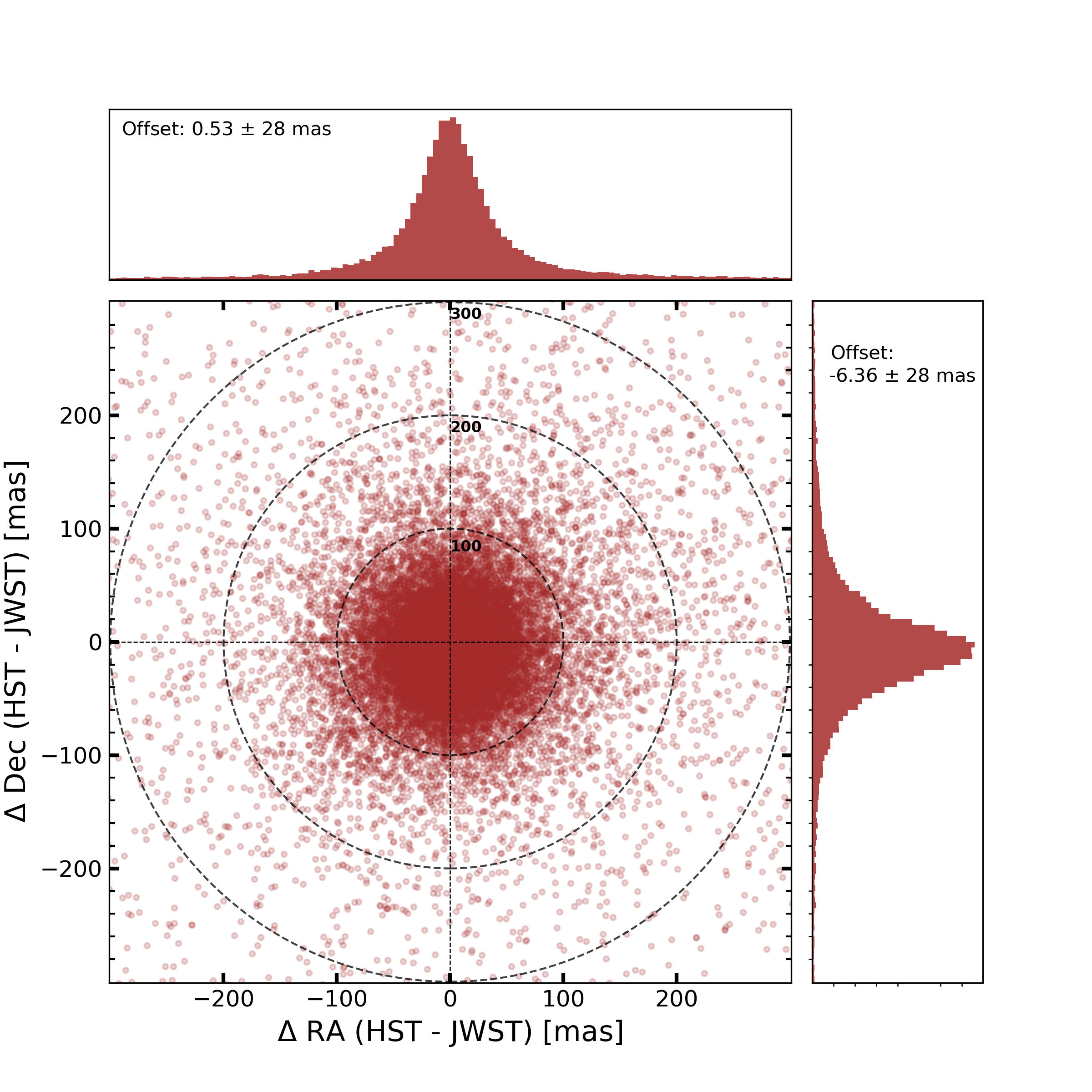}
	\caption{Astrometric offsets between MIRI F770W detections (S/N $\geqslant5$) and their HST/ACS F814W counterparts, which are aligned to Gaia DR3, assuming a cross-match radius of 1\arcsec. Shown for reference are \emph{dashed circles} with radii 0.1, 0.2, and 0.3$\arcsec$ centered on (0,0), depicting the relative positional offsets of all MIRI sources. The median offsets in RA and Dec.\ are 0.35 and 6 mas respectively (MIRI native pixel scale is 110 mas) with a median absolute deviation (MAD) of 28 mas in both directions.}
	\label{fig:astrometry}
\end{figure}

First is the astrometric correction performed by \emph{TweakReg}. Typically, this step attempts to align the astrometry to the Gaia DR3 astrometric reference frame \citep{Gaia2023}, however, the relatively small field-of-view of the MIRI imager ($\sim$2.3 arcmin$^2$) implies extremely few or almost zero matches with Gaia objects in each MIRI pointing. The \pipe, however, includes an option to align the astrometry based on a user-provided external reference catalog. For this purpose, a reference catalog was derived based on a mosaic constructed from an updated version of the COSMOS HST/F814W image \citep{Koekemoer2007}, covering the entire COSMOS-Web region (NIRCam and MIRI areas included). The astrometry of this HST mosaic was predominantly aligned to the Gaia DR3 reference frame \citep{Gaia2023} with some adjustments included to align with the COSMOS2020 catalog \citep{Weaver2022} as well as to ensure greater astrometric accuracy. Care was taken to exclude stars from this reference catalog that could bias the astrometry because due to proper motion. However, we note that there might be a faint population of field stars remaining in the catalog that might exhibit low-level bulk proper motion, whose proper motion is unknown if they are below the limits of the GAIA DR3 catalog.  In this step, we also generate our own source catalogs using \textsc{SE} for all MIRI exposures in a single visit, which are aligned relative to each other and eventually merged to create one combined source catalog. This combined catalog is then cross-matched and fit to the aforementioned COSMOS reference catalog, the results of which are propagated back to all the individual MIRI exposures, with the relative alignment between exposures remaining intact. Figure \ref{fig:astrometry} shows the differences in R.A. and Dec.~ between MIRI F770W and HST F814W for the cross-matched sources. The median offsets are 0.53 and 6 mas in R.A. and Dec. with a median absolute deviation (MAD) of 28 mas in each. We note that the MAD values are higher than the median offsets, which might be due to the low-level bulk proper motion of stars mentioned earlier, given the large time baseline between JWST and HST observations.

Even though we have already performed sky subtraction using \emph{SkyMatch} as an additional step in Stage 2, we run \emph{SkyMatch} again to normalize sky values with a median background level of zero for all MIRI exposures by setting \texttt{skymethod = `global+match'} and \texttt{subtract = True}. This is followed by the \emph{OutlierDetection} step -- where bad pixels or cosmic-ray hits missed in Stage 1 or other artifacts carried over from previous calibrations are flagged -- for which we use the default settings provided in the \pipe.

The final step is \emph{Resampling}, which resamples individual calibrated exposures and combines them to produce a single mosaic image. In order to overcome the computational challenges in handling the full NIRCam dataset, the COSMOS-Web region was divided into 20 smaller sub-regions, referred to as \textsf{tiles}, to streamline the processing. The dimensions and coordinates of these tiles are provided in the corresponding NIRCam data reduction paper (Franco et al., in prep.). These \textsf{tiles} have a fixed WCS solution that is aligned to the central coordinates of the COSMOS field (as adopted in previous works e.g., \citealt{Koekemoer2007,Laigle2016,Weaver2022}). The mosaics are also oriented 20$^{\circ}$ relative to North. The native pixel resolution of MIRI is 110 mas, and although producing mosaics with a 90 mas pixel scale would be sufficient, our mosaics are sampled at a finer resolution of 30 and 60 mas, matching the NIRCam mosaics, primarily to fulfill different scientific requirements. Alternatively, because several visits fall outside the coverage of our NIRCam data, we generate \textsf{individual} mosaic images for each MIRI visit, using the same setup applied to the \textsf{tile} mosaics.

\begin{table}
    \centering
    \begin{tabular}{ccc}
    \hline
    \# of Exposures & Area & Depth \\
     & (arcmin$^2$)& (AB)  \\
    \hline
    2 & 89.0 & 25.15\\
    4 & 436.9 & 25.51\\
    6 & 29.4 & 25.75\\
    8 & 127.8 & 25.85\\
    \hline
    \end{tabular}
    \caption{Survey depths as a function of number of exposures. The measured values are aperture corrected depths estimated using 0.27\arcsec radius apertures as discussed in Sec.\ \ref{sec:depth}.}
    \label{tab:depths}
\end{table}

\begin{figure*}[htp]
    \centering
    \includegraphics[width=2\columnwidth]{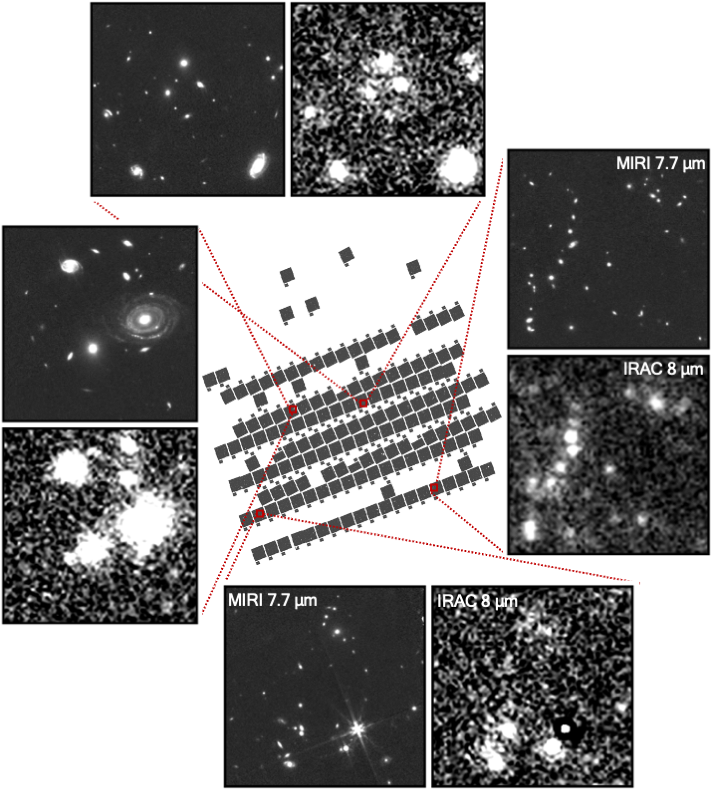}
    \caption{The COSMOS-Web MIRI F770W mosaic. The zoomed-in areas show 45\arcsec $\times$ 45\arcsec matched regions in both MIRI F770W and IRAC CH4 from different parts of the mosaic. These cutouts highlight the significant improvement in sensitivity ($\sim$$50\times$) and resolution ($\sim$$7\times$) of MIRI observations compared to IRAC. The montage of cutouts is shown with the same stretch for direct comparison.}
    \label{fig:MIRI_mosaic}
\end{figure*}

Figure \ref{fig:MIRI_mosaic} shows the full COSMOS-Web MIRI mosaic, including a comparison to IRAC CH4 from different areas of the mosaic. Our MIRI images are publicly available\footnote{\url{https://cosmos.astro.caltech.edu/page/cosmosweb-dr}} in two configurations: \textsf{tiles} (overlapping with the NIRCam-based tiles) and \textsf{individual} (all unique 160 MIRI visits). These are provided as FITS multi-extension format images known as \texttt{i2d} files containing the standard set of extensions\footnote{\url{https://jwst-pipeline.readthedocs.io/en/latest/jwst/data_products/science_products.html\#resampled-2-d-data-i2d-and-s2d}}, as well as individual images for the three main extensions: science (\textsf{SCI}), error (\textsf{ERR}), and weight (\textsf{WHT}) images.

\subsection{Survey Depth} \label{sec:depth}

The average 5$\sigma$ (point-source) depths as a function of the number of exposures are given in Table \ref{tab:depths}. Our depth estimates were derived using the \emph{ImageDepth} function of \texttt{PHOTUTILS} \citep{Bradley2023}. For each MIRI visit, we derive a \emph{source mask} by detecting sources based on a median background-subtracted convolved image (with a detection threshold of 3$\sigma$), and the image depth is then calculated by placing 200 random circular (non-overlapping) apertures on the blank regions of the image, which is derived based on the input \emph{source mask}. Here we adopt an aperture radius equal to the full width at half maximum (FWHM) of the F770W point spread function (PSF) i.e., 0.27\arcsec, based on in-flight measurements from commissioning data \footnote{\url{https://jwst-docs.stsci.edu/jwst-mid-infrared-instrument/miri-performance/miri-point-spread-functions\#gsc.tab=0}; See Table 1}. Aperture corrections are applied based on empirical PSF estimates from \citet{Libralato2024} as discussed in Section \ref{sec:phot}. As reported by \citet{Casey2023}, the measured depths are significantly deeper than the exposure time calculator (ETC; v2.0) predictions, with differences ranging from $\sim0.7-0.8$ mag for the different number of exposures.

\section{Source Catalog} \label{sec:source_cat}

The MIRI F770W photometric measurements were derived in two different ways to serve different scientific goals. Model-based photometry using empirical F770W PSFs (derived using PSFex, \citealt{Bertin2011}) as a part of the multi-wavelength catalog based on COSMOS-Web and overlapping archival ground- and space-based observations is provided in the COSMOS-Web catalog (Shuntov et al., in prep.). This catalog is primarily based on source detection using deep NIRCam imaging from COSMOS-Web, with photometry derived using the light profile-fitting method in \textsc{SourceXtractor++} (SE++; \citealt{Bertin2020}) for imaging data in 34 photometric bands in the COSMOS field. In this work, however, we use F770W photometry from the COSMOS-Web MIRI catalog that is derived using \textsc{SE} based on just the F770W imaging, as discussed in the following text. Since this catalog is primarily based on source detection in the MIR, it is also used to find potential MIRI-only sources in COSMOS-Web, which will be explored in future work.

\subsection{Source Extraction and Photometry} \label{sec:phot}
We performed source detection and photometry using \textsc{SE} with the main parameter settings listed in Table \ref{tab:sextractor_params}, while adopting default values for the remaining parameters. In order to limit the number of spurious objects, we masked all bright, saturated sources (stars and galaxies) with diffraction spikes in each MIRI visit. Subsequently, we ran \textsc{SE} in dual-image mode with the masked image as the detection image with photometry measured using the original MIRI image. We adopt the photometry given by \texttt{FLUX\_AUTO} as the total measured flux in our source catalog. These flux measurements are based on an adaptively scaled aperture, commonly known as \emph{Kron} photometry, which are typically elliptical apertures with a certain scaling factor called the \emph{Kron} parameter (\emph{K}). As shown in Table \ref{tab:sextractor_params}, we set $K = 2.5$ with a minimum \emph{Kron} radius value set to 2.0. For compact and/or faint objects, especially those in higher background regions, \emph{Kron} photometry can be unreliable due to artificially inflated apertures \citep{Graham2005}. Therefore, we also measured photometry using fixed (circular) apertures with diameters 0.3, 0.6, 1.0, 1.5 and 2.0\,$\arcsec$, which are mainly useful for point sources.

\begin{table}
\begin{threeparttable}
\centering
    \noindent\begin{tabular}{lc}
        \hline
        \hline
        \textbf{Extraction}\\
        \texttt{DETECT\_MINAREA} & 10 pixels\\
        \texttt{DETECT\_THRESH} & 1.5$\sigma$\\
        \texttt{ANALYSIS\_THRESH} & 1.5$\sigma$\\
        \texttt{FILTER\_NAME} & gauss\_3.0\_5x5.conv\\
        \hline
        \textbf{Photometry}\\
        \texttt{PHOT\_AUTOPARAMS} & 2.5, 2.0\\
        \texttt{MAG\_ZEROPOINT} & 26.581$^a$\\
        \texttt{PIXEL\_SCALE} & 0.06 \\
        \hline
        \textbf{Star/Galaxy Separation}\\
        \texttt{SEEING\_FWHM} & 0.27\arcsec\ $^b$ \\
        \hline
        \textbf{Background}\\
        \texttt{BACK\_TYPE} & AUTO \\
        \texttt{BACK\_SIZE} & 32 \\
        \texttt{BACK\_FILTERSIZE} & 3 \\
        \texttt{BACKPHOTO\_TYPE} & LOCAL \\
        \hline
        \textbf{Deblending}\\
        \texttt{DEBLEND\_NTHRESH} & 32 \\
        \texttt{DEBLEND\_MINCONT} & 0.008 \\
        \hline
        \hline
    \end{tabular}
    \caption{Main \textsc{SE} configuration parameters used for MIRI F770W photometry in the COSMOS-Web MIRI catalog. All other parameters are set to program defaults.}
    \begin{tablenotes}
        \small{
        \item $^a$ The magnitude zeropoint (in units of AB) of images, converting from the JWST default of MJy sr$^{-1}$ to $\mu$Jy pixel$^{-1}$ for a \texttt{PIXEL\_SCALE} of 0.06\arcsec.
        \item $^b$ \texttt{SEEING\_FWHM} is set to the PSF FWHM of F770W based on in-flight measurements.
        }
    \end{tablenotes}
    \label{tab:sextractor_params}
\end{threeparttable}
\end{table}

\begin{figure}
    \centering
    \hspace{-3em}\includegraphics[width=1.1\linewidth]{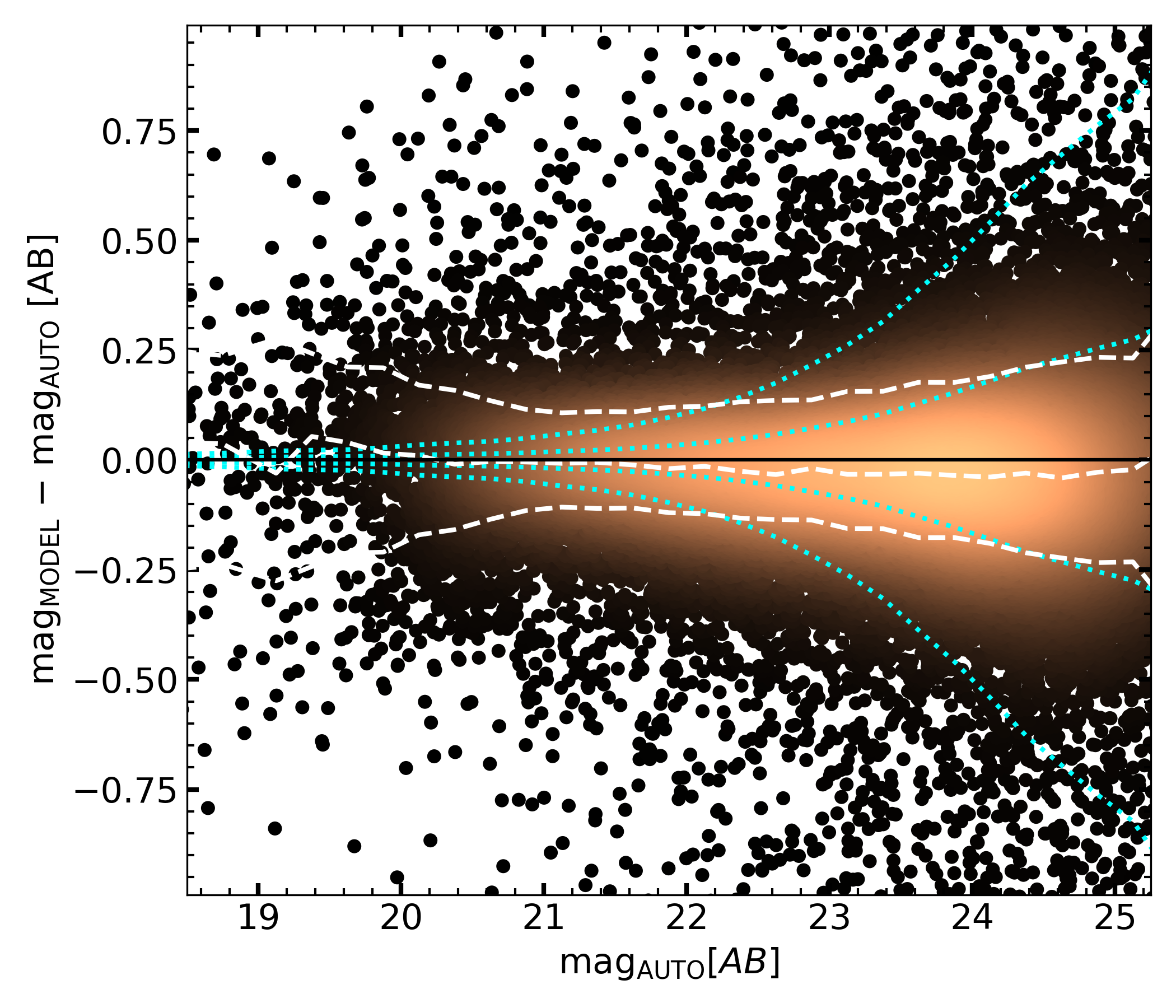}
    \caption{Photometric comparison between SE++ model magnitudes and SE AUTO aperture magnitudes for MIRI F770W. The median offset along with the 3$\sigma$-clipped standard deviation is shown by the \emph{white dashed lines}. Also shown for reference are the $\pm1\sigma$ and $\pm3\sigma$ photometric uncertainties (\emph{cyan dotted lines}) combined in quadrature from the two photometric measurements.}
    \label{fig:sep_se_comp}
\end{figure}

\textsc{SE} derived photometric uncertainties typically account for photon and detector noise only, but not the pixel-to-pixel correlated noise. In order to account for this uncertainty, we derive an empirical noise estimate based on the random empty aperture method. Similar to the procedure mentioned in Section \ref{sec:depth}, we placed 1500 random circular apertures with aperture diameters ranging $0.3 - 3\arcsec$ across each MIRI visit (using noise-equalized images). For each aperture diameter, we measured the noise by fitting a Gaussian to the negative side of the flux distribution, thereby avoiding any flux from real sources. Since the measured noise is dependent on the number of pixels in a given aperture, we fit a power law to this distribution following \citep{Finkelstein2022} and \citep{Rieke2023} as follows: 
\begin{equation}
    \sigma_N = \sigma_1\alpha N^\beta,
\end{equation}
where $\sigma_N$ is the noise in an aperture with $N$ pixels, and $\sigma_1$ is the pixel-to-pixel noise measured per image as the sigma-clipped standard deviation of all pixels excluding source detections, and $\alpha$, $\beta$ are free parameters which we fit for using Python's \texttt{curve\_fit} function. Based on these functional form fits, we derive the total photometric uncertainties for each source in the catalog using the number of pixels in the respective \emph{Kron} aperture (area = $\pi \times$ A\_IMAGE $\times$ B\_IMAGE $\times$ KRON\_RADIUS$^2$, where $A$ and $B$ are the semi-major and semi-minor axes of the elliptical apertures), with the \textsc{SE} measured uncertainties added in quadrature. 

For both Kron and fixed-aperture photometry, we calculate aperture corrections based on empirical PSF (ePSF) from \citet{Libralato2024} derived using MIRI imaging from multiple publicly available JWST (Commissioning, Cycle 1 and GO) programs. This PSF version takes into account effects such as the cruciform artifact seen in the bluest MIRI filters, F560W and F770W. We compute the encircled flux (EEF) of the ePSF over different radii, and the aperture corrections are derived as 1/EEF(r), where \emph{r} is the circular aperture radius for fixed-aperture photometry, and $r$ = KRON\_RADIUS $\times \sqrt{A\_IMAGE\times B\_IMAGE}$ in the case of Kron apertures.

\subsection{Comparison with model-based photometric catalog}\label{sec:se_sep_comp}
Figure \ref{fig:sep_se_comp} compares the AUTO photometry for F770W from the COSMOS-Web MIRI catalog with the model-based SE++ photometry from the COSMOS-Web galaxy catalog. Only sources that were cross-matched between both catalogs were included in this comparison. Since the AUTO fluxes are measured using \emph{Kron}-like elliptical apertures that scale according to the brightness of the sources and include PSF-based aperture corrections, they are a good proxy for the total flux estimates and therefore a fair comparison with the SE++ model fluxes. Overall, there is good agreement between the two measurements with a running median offset (mag$_{MODEL}$ -- mag$_{AUTO}$) between -0.04 and 0.06 mag with a median scatter of $\sim$0.22 mag. Due to the paucity of sources on the bright end, the median offset as well as the scatter is slightly higher; however, for the rest of the source distribution, the observed median offset shows that these measurements are largely consistent within the $\pm1\sigma$ combined photometric uncertainties. 

\begin{figure}[]
	\centering
	\hspace{-3em}\includegraphics[width=1.1\linewidth]{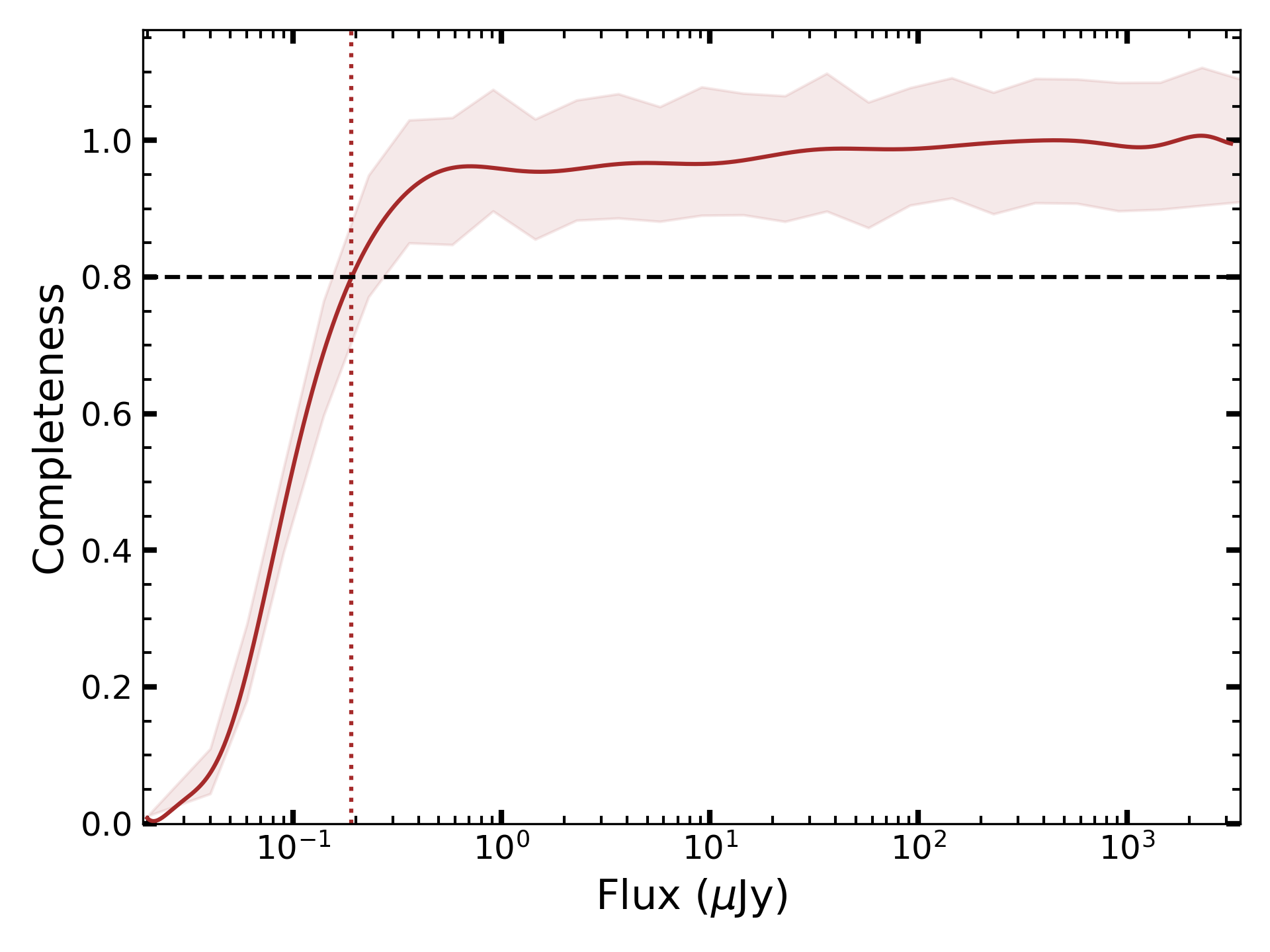}
	\caption{Completeness fraction as a function of MIRI F770W flux. The uncertainties (\emph{shaded region}) shown here are purely Poisson errors. The 80\% completeness limit (\emph{horizontal dashed line}) as well as the average 5$\sigma$ limiting depth (\emph{vertical dotted line}) are shown for reference. Our MIRI source sample achieves $\geqslant 80$\% completeness for all sources detected with SNR $\geqslant 5\sigma$.}
	\label{fig:completeness}
\end{figure}

\begin{figure*}
    \centering
    \includegraphics[width=1\linewidth]{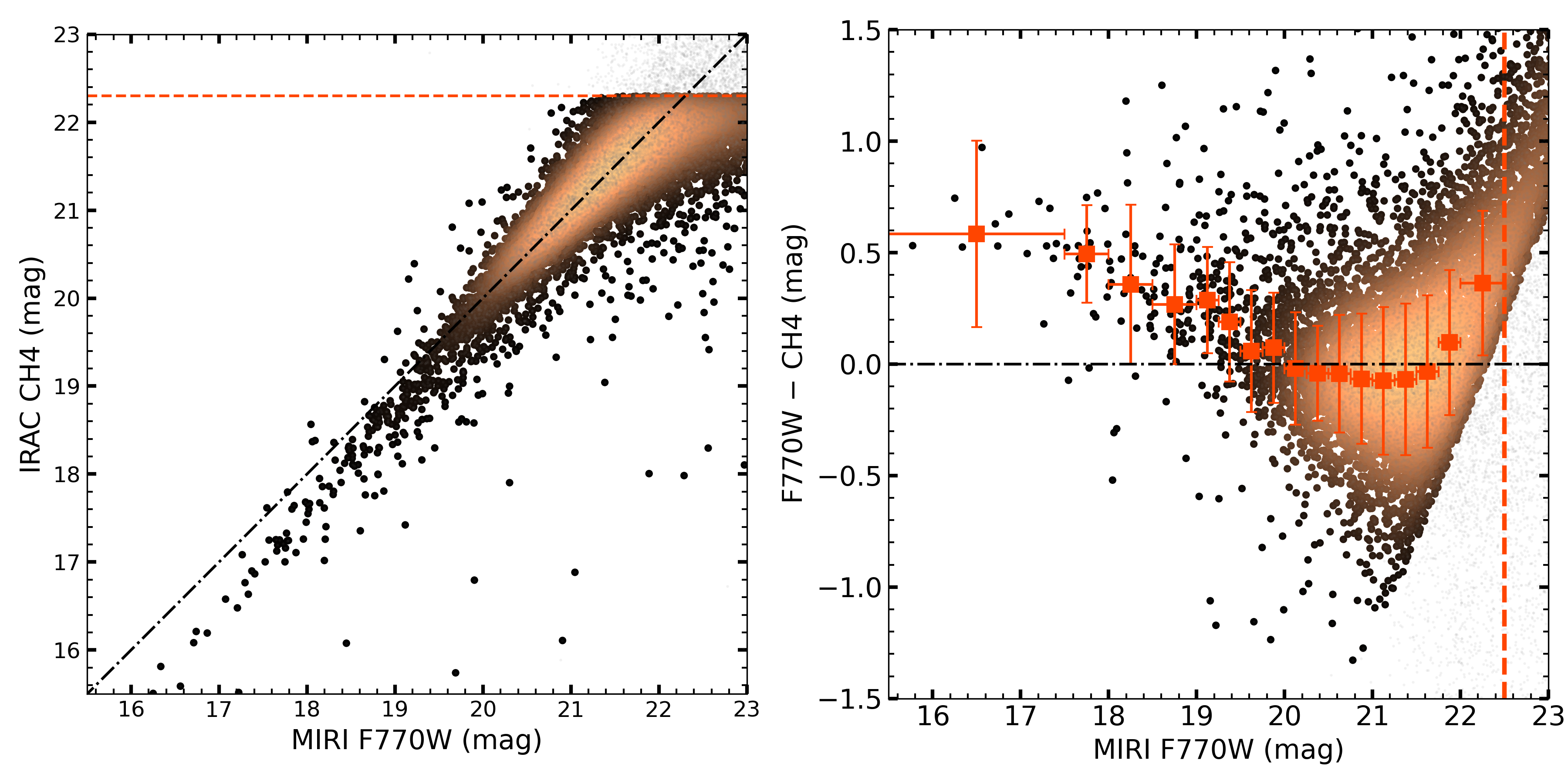}
    \caption{Photometric comparison of MIRI F770W detections with IRAC CH4 measurements from the COSMOS2020 catalog. For the sake of accuracy, we compare MIRI photometry with IRAC CH4 for only sources with SNR$_{CH4} \geqslant 5$ and \emph{mag}$_{CH4} \leqslant 22.3$ in the COSMOS2020 catalog(\emph{brown-black points}). \emph{Left} panel shows a 1:1 comparison with the identity line (\emph{diagonal dashdot line}) and the $5\sigma$ limiting depth for IRAC CH4 (\emph{horizontal dotted line}) shown for reference. For the matched sample, the magnitude offsets ($\Delta_{mag}$ = F770W$-$ CH4) as a function of the MIRI F770W photometry is shown in the \emph{right} panel. Also shown are median offsets per magnitude bin (\emph{orangered squares}) derived by binning sources based on their MIRI F770W photometry along with their 3$\sigma$-clipped standard deviation. Some of the bins at the bright end ($m_{F770W} < 19.5$) have asymmetric widths because we required each bin to contain a minimum of 10 sources.}
    \label{fig:miri_irac_comp}
\end{figure*}

\subsection{Completeness Simulations} \label{sec:compl}
A measure of source completeness -- the fraction of objects detected at a given flux density given the image noise properties -- is imperative to accurately estimate the statistical properties of a given sample. For this purpose, we employ the \texttt{ComEst}\footnote[2]{\url{https://github.com/inonchiu/ComEst}} \citep{Chiu2016} package to derive the source completeness in our images. The fundamental idea behind \texttt{ComEst} is to simulate synthetic point sources and galaxies on the observed ``source-free" image using the image simulation toolkit, \textsc{GalSim} \citep{Rowe2015}, and then run \textsc{SE} on the same image to derive the source detection rate as a function of flux (or magnitude). Initially, \texttt{ComEst} runs \texttt{SE} to generate a first-order source catalog along with various check-images, and thereafter removes the detected sources from the observed image and replaces those pixel values by randomly sampling the background values (assuming a normal distribution with mean and standard deviation in the \textsf{BACKGROUND} and \textsf{BACKGROUND\_RMS} check-images provided by \texttt{SE}), thereby producing the \emph{source-free image} (SFI). This is followed by the injection of `fake' sources, simulated using \texttt{GalSim}, on the SFI to create a set of simulated images on which \texttt{SE} is again run to detect the simulated sources. Finally, the completeness as a function of source flux (or magnitude) is derived by comparing \texttt{SE} catalogs of simulated images with the \emph{true} catalog that is used as an input to the simulation. 

For source injection on the SFI, we chose to simulate only point sources since a majority of the objects that are resolved in F770W are either bright or extended or both (e.g., the half-light radius, $r_{50}$, for $\sim80$\% of the sources with $m_{F770W}<21$ is greater than the FWHM of the F770W PSF) while most of the faint sources -- the regime of higher source incompleteness -- are typically unresolved (or, marginally resolved at best) and appear compact in our observations. For simulation of point sources, \texttt{ComEst} simply convolves each source with a Moffat PSF ($\beta=4.5$) using the user-defined FWHM, which in this case is equal to the F770W's FWHM value of 0.27$\arcsec$. The number density of the simulated sources per arcmin$^2$ is set to 30, and the source flux distribution ranges from 0.01 to 2500 $\mu$Jy. \texttt{ComEst} can simulate multiple images with the same configuration scheme in parallel, so we generate 20 simulated images for each unique MIRI visit in our program, and \texttt{ComEst} eventually produces a merged catalog each for both simulated images \texttt{SE} catalogs and the input \emph{true} catalog for each simulation. The resulting completeness estimates as a function of source flux are shown in Figure \ref{fig:completeness}. As shown in the figure, we achieve 80\% completeness corresponding to the median 5$\sigma$ depths measured in our images, i.e., $\sim$0.2 $\mu$Jy.

\section{Results and Discussion} \label{sec:results}

\subsection{Comparison of JWST/MIRI and Spitzer/IRAC photometry at 7.7\,$\mu$m} \label{sec:miri_irac_comp}
We compare our MIRI F770W fluxes with the IRAC CH4 (8\,$\mu$m) fluxes from the COSMOS2020 catalog \citep{Weaver2022} as shown in Figure \ref{fig:miri_irac_comp}. The reported 5$\sigma$ depth for IRAC CH4  is 22.5 mag, so we restrict our comparison to only source detections with SNR$_{CH4} \geqslant 5$ and $m_{CH4} \leqslant 22.5$ mag. Based on the binned median offsets ($\Delta_{\textnormal{mag}} = m_{F770W} - m_{CH4}$), the overall agreement is reasonable with offsets that are fairly small ($\Delta_{\textnormal{mag}} \leqslant 0.05$ mag) with a 3$\sigma$-clipped standard deviation of 0.26 mag for the bulk of the distribution between $19.5 < m_{\textnormal{F770W}} < 22$. Previous studies \citep[e.g.][]{Yang2023b} have shown that this level of offset ($<0.1$ mag) is reasonable, especially given the differences in filter transmission for MIRI and IRAC and also the fact that these observations are calibrated independently. However, the offsets become larger towards the faint end ($m_{F770W} > 22$), as we approach the limiting depth for CH4, which might be partly driven by large uncertainties in the CH4 photometry. At the bright end, $m_{F770W} < 19.5$, there exists a systematic brightening of the CH4 photometry ($\Delta_{\textnormal{mag}} > 0.2$) compared to F770W. In the following text, we postulate probable reasons for the observed offsets between the two photometric measurements.

The IRAC CH4 photometry from the COSMOS2020 catalog \citep{Weaver2022} is model-based, which is jointly derived using the $izYJHK_s$ imaging at optical/NIR wavelengths, where IRAC fluxes are obtained by convolving the model with the spatially varying CH4 PSF across the entire coverage. If the source properties such as the size or shape vary significantly as a function of wavelength, then this might result in uncertain flux measurements, particularly for objects that are faint in the modeled bands but bright in the MIR. We note that most of our bright sources ($m_{F770W} < 19.5$) are at low redshifts, $z<0.5$, which might result in systematic offsets for objects at a given brightness due to morphological differences between the modeled bands and CH4. In contrast, the F770W photometry is directly based on high-resolution images at MIR wavelengths. Also, because of an order of magnitude broader PSF in the case of CH4, the IRAC photometry is significantly affected by source blending compared to MIRI, especially for bright sources. This effect has also been observed in other studies comparing MIRI and IRAC photometry with F770W \citep{Yang2023b} and F560W \citep{Sajkov2024,Ostlin2024} filters.

\begin{figure*}
    \centering
    \includegraphics[width=\linewidth]{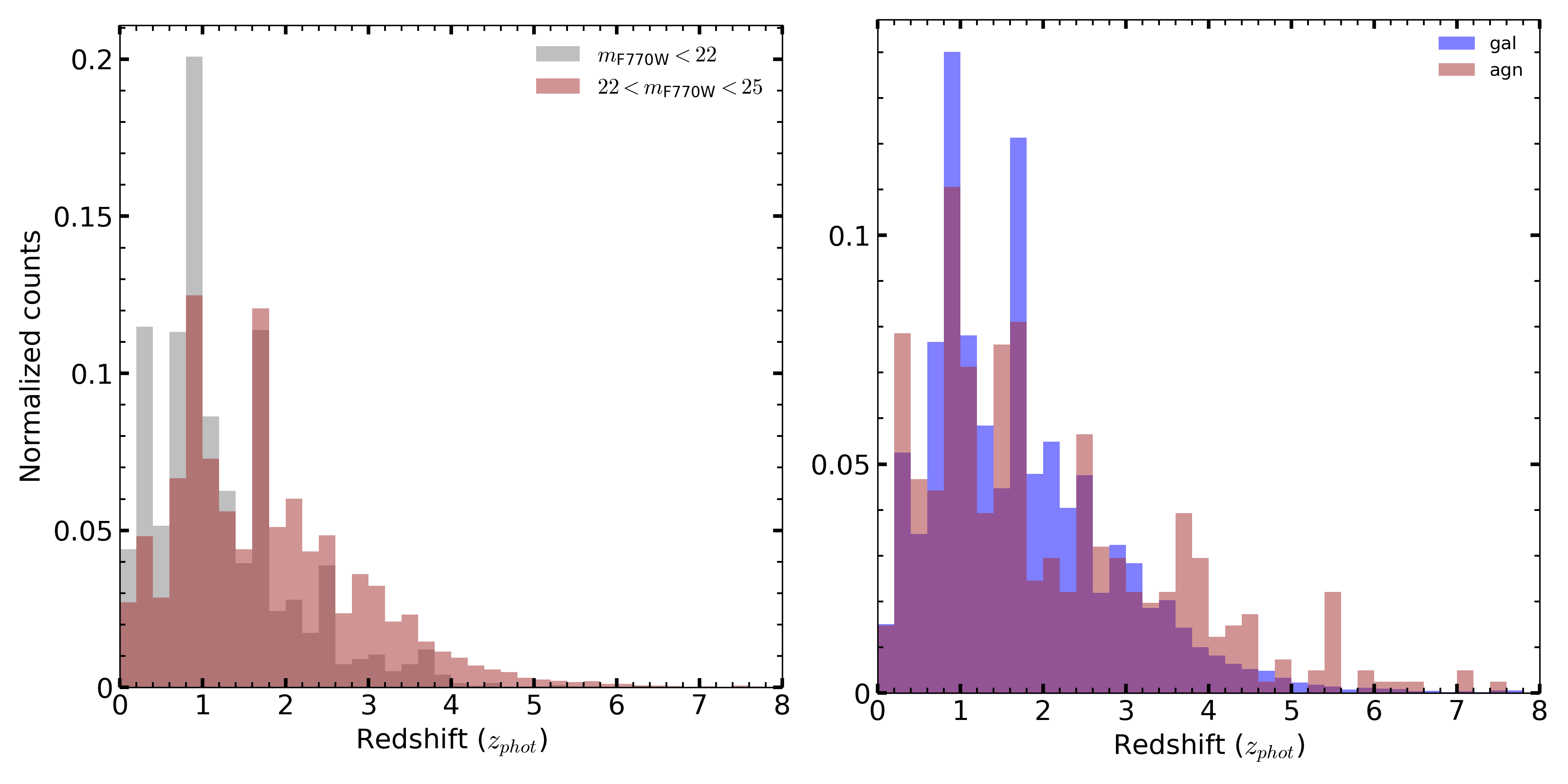}
    \caption{Photometric redshift distribution of the 7.7$\mu$m MIRI-selected sources from the COSMOS-Web catalog. \emph{Left}: Redshift distribution for the sub-samples based on two different brightness thresholds. The bright end ($m_{F770W} < 22$) is dominated by low redshift sources ($z<2$) whereas higher redshift sources ($z>2$) become more prevalent as we move towards the faint end. \emph{Right}: Redshift distribution for the sub-samples based on the source classification from LePHARE measurements.}
    \label{fig:redshift_dist}
\end{figure*}

\subsection{Photometric redshifts}
Using a modified version of the template-fitting code LePHARE \citep{Arnouts2002,Ilbert2006}, photometric redshifts are provided in the COSMOS-Web catalog based on multi-band photometry including MIRI F770W where available. Compared to previous versions of the COSMOS catalog \citep{Laigle2016,Weaver2022}, the current version of redshift fits is based on a diverse and representative library of galaxy templates (stellar and AGN) as well as a wide range of dust attenuation parameters, especially with the inclusion of F770W imaging where PAH emissions from low redshift sources could contribute significantly to the galaxy's SED. For more information on the quality of photo-z fits and their performance with respect to the spectroscopic redshifts, we refer the reader to our COSMOS-Web catalog paper (Shuntov et al. \emph{in prep}). 

Figure \ref{fig:redshift_dist} shows the redshift distribution of the 7.7$\mu$m MIRI-selected sources. The distribution spans a wide redshift range from $z\sim0$ through $z\sim8$ with $\sim80\%$ of the sources lying at $z\leqslant3$. Sources with $z>4$ constitute $\sim5\%$ of our sample. As seen in Figure \ref{fig:redshift_dist}, the majority of bright sources ($m_{F770W} < 22$) are dominated by low redshift objects with $z\leqslant2$, whereas the higher redshift sources are more prevalent at fainter magnitudes as expected. The defined peaks in the distribution at $z\sim 1$$-$2 are due to galaxies with increasing stellar continuum emission dominating the 7.7$\mu$m wavelengths and potentially the broad 3.3$\mu$m PAH emission contributions for star-forming galaxies at these redshifts. Using the best-fit $\chi^2$ values for various templates, the $Bzk$ color criteria \citep{Daddi2004}, and source properties such as the effective radius $R_{\textnormal{eff}}$ derived the LePHARE measurements, sources in the COSMOS-Web catalog are classified into stars, galaxies and AGN. For the MIRI-selected sources, we find a diverse population of galaxies and AGN spanning a wide redshift range, with AGN mostly dominating at $z\geqslant3.5$. However, we note that the AGN classification is somewhat less robust. For a more detailed and accurate determination of photo-$z$ for AGN, we refer interested readers to the work by \citet{Salvato2011}.

\subsection{Number Counts}\label{sec:counts}

\begin{figure*}
	\hspace{-1em}\includegraphics[width=2.2\columnwidth]{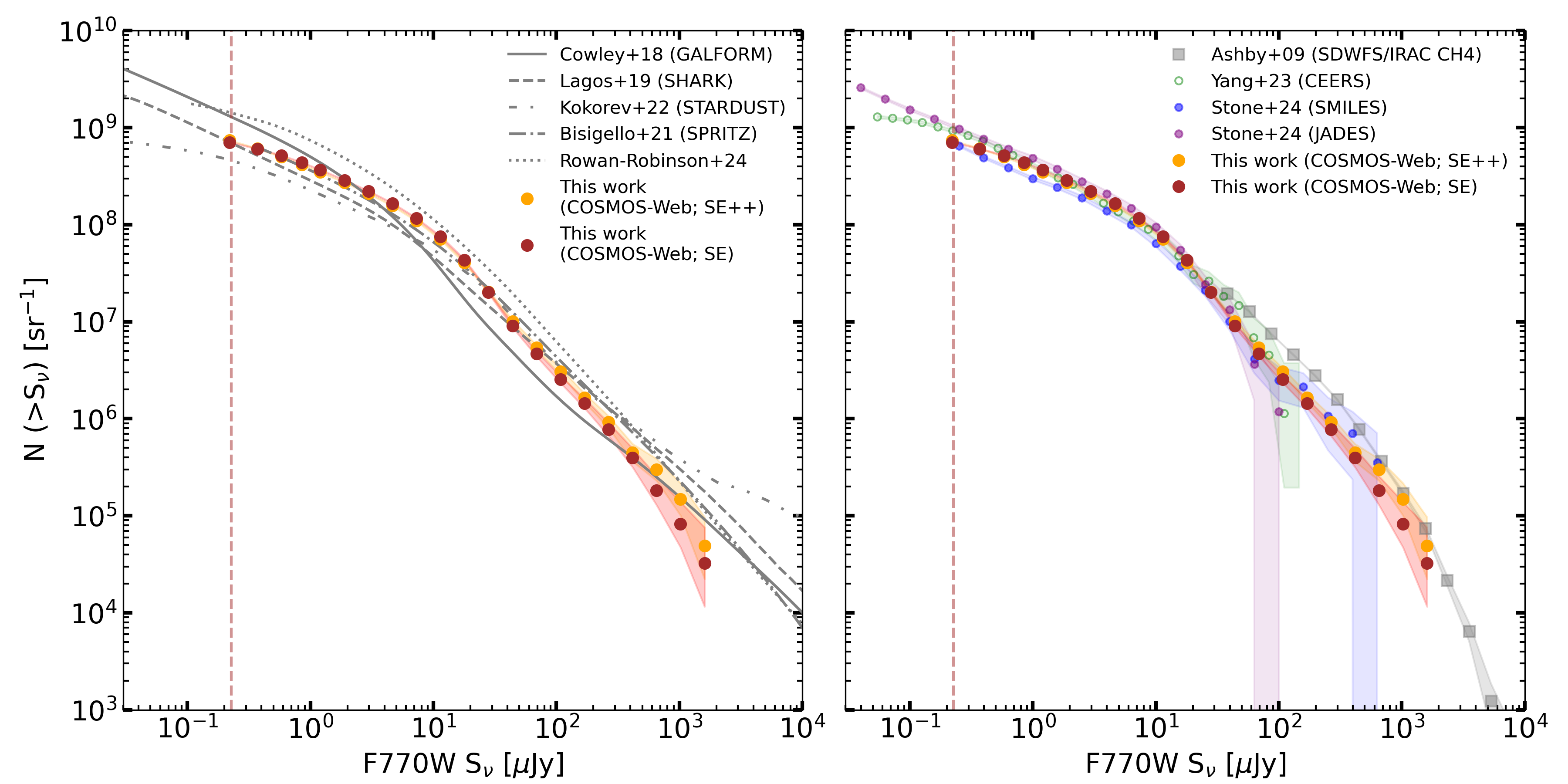}
	\caption{Cumulative number counts at 7.7$\mu$ from COSOMS-Web in comparison with model predictions (\emph{left}) and other JWST and IRAC number counts from literature (\emph{right}). Uncertainties for the COSMOS-Web number counts are a combination of Poisson and cosmic variance errors added in quadrature. The \emph{dashed vertical line} in both panels represent the average 5$\sigma$ limiting depth achieved in the COSMOS-Web MIRI observations, which also correspond to the 80\% completeness limit as mentioned in Section \ref{sec:compl}. In the \emph{left} panel, we compare our source counts with a number of model predictions at 8$\mu$m including those based on GALFORM+GRASIL \citep{Cowley2018} and SHARK \citep{Lagos2019} semi-analytical models, the SPRITZ \citep{Bisigello2021} simulations, and the semi-empirical models of \citet{Rowan-Robinson2009,Rowan-Robinson2024} and \citet{Kokorev2021}. We also compare our number counts with estimates from recent JWST (MIRI F770W) studies including the CEERS \citep{Yang2023b}, JADES and SMILES \citep{Stone2024} programs, as well as \emph{Spitzer} (IRAC CH4) counts from the SDWFS \citep{Ashby2009} program.}
	\label{fig:num_counts}
\end{figure*}

We derive the MIRI 7.7$\mu$m cumulative number counts (completeness corrected) based on the COSMOS-Web MIRI catalog described in Section \ref{sec:source_cat}. Given that we consider only sources in flux bins with completeness fractions $\geqslant 80\%$ towards number counts estimation, $\sim85\%$ of the total MIRI coverage reaches imaging depths that are at least $\sim$25.5 mag or better, which is comparable to the average 5$\sigma$ limiting depths reported in Section \ref{sec:depth}. Therefore, we assume that our coverage is largely uniform across visits, so the total effective area as a function of flux bins remains the same. Figure \ref{fig:num_counts} shows our 7.7$\mu$m number counts that are completeness corrected based on the simulations presented in Section \ref{sec:compl}, along with Poisson errors that purely reflect statistical uncertainties. The corresponding values are provided in Table \ref{tab:numcounts}. 

For comparison, the 7.7$\mu$m number counts based on the model-based photometry from the COSMOS-Web catalog is also shown in Figure \ref{fig:num_counts}, which is overall in good agreement with those based on the COSMOS-Web MIRI catalog (within 1$\sigma$ uncertainties), except for a small but negligible overprediction on the bright end;  this might be partly due to the larger offsets observed between the two photometric measurements for bright sources, as discussed in Section \ref{sec:se_sep_comp}.

\subsubsection{Cosmic Variance}
In most extragalactic surveys, especially with smaller coverage, number or density measurements derived from galaxy populations as a whole are prone to cosmic variance uncertainties since the Universe is inhomogeneous on small scales \citep{Driver2010}. The impact of cosmic variance is typically lower for surveys with larger sky coverage and those including multiple distinct sightlines. Also, cosmic variance is higher for massive high-redshift populations and brighter sources in general \citep{Moster2011}.

Since COSMOS-Web has a large areal coverage, our reported 7.7$\mu$m number counts will be minimally impacted by cosmic variance compared to similar measurements from smaller area JWST surveys (e.g., JADES and SMILES). Given that it is a contiguous field, the cosmic variance uncertainties might be significant on the bright end of the number counts, however. As discussed in Sec. \ref{sec:obs_counts}, the small offset observed between IRAC and MIRI number counts could be partially due to cosmic variance among other things such as filter differences (Sec. \ref{sec:miri_irac_comp}) and the presence of stars in the IRAC estimates. Using the empirical expression given by \citet{Driver2010}, we estimate the potential cosmic variance\footnote{\url{https://cosmocalc.icrar.org/}} impact on the observed COSMOS-Web 7.7$\mu$m number counts as follows. Based on the observed redshift distribution, more than 70\% of the sources are at redshifts $z \leqslant 3$; since the number counts are dominated by galaxies at these redshifts, we derive the cosmic variance uncertainties assuming a source distribution with $0.01 < z < 3$. For a total area of $\sim0.2$ deg$^2$ (assuming cosmological parameters based on \citealt{Planck2018}), the expected cosmic variance is $\sim 6.5$\%, which is added in quadrature to the Poisson errors mentioned in Section \ref{sec:counts}. 

\subsubsection{Comparison with other observations}\label{sec:obs_counts}
As mentioned earlier, a number of recent studies have derived number counts at different mid-IR wavelengths (5.6-25.5 $\mu$m) using MIRI observations from various JWST programs \citep[e.g.,][]{Ling2022,Wu2023,Yang2023b,Stone2024,Sajkov2024}. Figure \ref{fig:num_counts} (\emph{right}) includes a comparison of our number counts at 7.7$\mu$m with corresponding estimates from the (1) CEERS program given by \citet{Yang2023b}, (2) SMILES and JADES program given by \citet{Stone2024}. Overall, up to the 80\% completeness flux limits, there exists a good overlap with these other estimates within the 1$\sigma$ uncertainties. On the faint end ($S_\nu \lesssim 1 \mu$Jy), a small offset exists between estimates based on SMILES observations and other programs that is attributed to potential cosmic variance effects as reported by \citet{Stone2024}, which is the likely cause of typical variance in observed counts between different surveys. However, towards the bright end ($S_\nu \gtrsim 20 \mu$Jy), there exists small differences between various estimates. This is mostly driven by the fact that most of these other programs probe significantly smaller areas ($\sim$ 10-35 arcmin$^2$), thereby resulting in large scatter in the brightest flux bins. The power of COSMOS-Web really shines in this regime thanks to the large areal coverage of $\sim$ 0.2 deg$^2$, which is 30-70 times the area observed by other programs.

We also compare our number counts to those based on IRAC CH4 detections from the Spitzer Deep Wide Field Survey (SDWFS; \citealt{Ashby2009}), which observed a large $\sim$10 deg$^2$ in the Bootes field with an average 5$\sigma$ limiting depth of 20.18 mag ($\sim$30 $\mu$Jy). We derived the CH4 number counts based on the 8$\mu$m source flux measurements from the 8 $\mu$m-selected catalog provided by \citet{Ashby2009}. As shown in Figure \ref{fig:num_counts} (\emph{right}), there is a good agreement between estimates on the bright end of this comparison (within 1$\sigma$ uncertainties); however, the CH4-based number counts are systematically higher compared to the MIRI-based estimates except for a limited overlap with CEERS and SMILES. The IRAC CH4 filter bandwidth is slightly higher and extends to slightly longer wavelengths ($\sim$9.3 $\mu$m) compared to F770W ($\sim$8.7 $\mu$m). Therefore, in the case of a small fraction of sources at low redshifts, the presence of a bright PAH feature that is redshifted out of F770W's bandpass, but still detectable in CH4, will result in slightly brighter fluxes in CH4 compared to F770W (e.g., see Fig.~7 in \citealt{Yang2023b}); however, this is expected to have minimal effect on the measured number counts when comparing MIRI to IRAC counts. In summary, our measured number counts are consistent with counts derived from other JWST observations (within 1$\sigma$ uncertainties) but slightly underpredict compared to counts based on Spitzer observations. COSMOS-Web is the only JWST-based survey to date to have efficiently sampled a wide flux range over a large area, thereby providing a comprehensive estimate of the number counts at 7.7$\mu$m. 

\begin{table}
    \centering
    \caption{Cumulative Number Counts at 7.7$\mu$m}
    \begin{tabular}{cc}
    \hline\hline
    Flux & N($>$S$_{\nu}$)\footnote{The reported counts are based on the COSMOS-Web MIRI catalog which include completeness corrections and the reported errors include Poisson as well cosmic variance uncertainties.} \\
    ($\mu$Jy) & (10$^{-6}$sr$^{-1}$)\\
    \hline
        0.22 & 705.76 $\pm $3.4\\
        0.37 &  605.72 $\pm$ 3.2\\
        0.58 &  517.079 $\pm$ 2.9\\
        0.85 &  437.75 $\pm$ 2.7\\
        1.21 &  366.04 $\pm$ 2.5\\
        1.89 &  287.30 $\pm$ 2.2\\
        2.96 &  221.30 $\pm$ 1.9\\
        4.64 &  165.63 $\pm$ 1.7\\
        7.28 &  116.06 $\pm$ 1.4\\
        11.41 &  75.30 $\pm$ 1.1\\
        17.88 &  43.29 $\pm$ 0.8\\
        28.03 &  20.21 $\pm$ 0.6\\
        43.94 &  9.12 $\pm$ 0.4\\
        68.88 &  4.67 $\pm$ 0.3\\
        107.98 &  2.54 $\pm$ 0.2\\
        169.27 &  1.44 $\pm$ 0.2\\
        265.34 &  0.78 $\pm$ 0.1\\
        415.96 &  0.40 $\pm$ 0.1\\
        652.06 &  0.18 $\pm$ 0.1\\
        1022.17 &  0.08 $\pm$ 0.04\\
        1602.37 &  0.03 $\pm$ 0.02\\
    \hline
    \end{tabular}
    \label{tab:numcounts}
\end{table}

\subsubsection{Comparison with models}

Typically, it has been challenging for galaxy simulations to reproduce the observed number counts \citep[e.g., see][]{Somerville2012,Lacey2016} since galaxies with different masses and from different cosmic epochs contribute to this observable, and it is hard for simulations to predict accurately the expected galaxy counts across the entire history of the universe and also over a wide wavelength range (e.g., far-UV to far-IR). Therefore, reconciling the observed number counts with predictions from theoretical and/or semi-analytical models of galaxy formation and evolution is an important exercise for improving our modelling efforts.  Figure \ref{fig:num_counts} (\emph{left}) shows number count predictions based on several different models in comparison with our observed counts. In the following text, we provide a brief overview of these different models along with a discussion on the conformity of the observed counts with model predictions.

The \citet{Cowley2018} predictions are based on a model combining a dark matter only \emph{N}-body simulation, the semi-analytical model, \textsc{GALFORM} \citep{Cole2000}, and the dust radiative transfer models from \textsc{GRASIL} \citep{Silva1998}, which account for interstellar dust absorption and re-emission of stellar radiation in computing the galaxy SEDs. Apart from the fiducial model, they also provide another version that includes evolving SN feedback as a function of redshift, which is able to better match the reionization redshift as inferred from the Planck cosmic microwave background data \citep{Planck2016} compared to the fiducial model. In this work, we compare our counts with the ``evolving feedback'' model that is in better agreement with observational data from both low and high redshifts. 

The \citet{Lagos2019} predictions utilize a combination of the semi-analytical model, \textsc{SHARK} \citep{Lagos2018}, and radiative transfer estimates from the EAGLE simulations \citep{Trayford2017} for dust attenuation along with Spitzer-based IR templates \citep{Dale2014} to account for dust re-emission. These predictions have achieved considerable success in reproducing the observed number counts from UV to the FIR (assuming a universal IMF) owing to accurate estimates of galaxy properties including, but not limited to, dust masses and temperature, gas metallicites, galaxy sizes and gas content, to name a few.

Using a novel panchromatic SED fitting tool, \textsc{Stardust}, \citet{Kokorev2021} provide a multi-parameter catalog containing infrared properties of $\sim$5000 star-forming galaxies from the Great Observatories Origins Deep Survey North (GOODS-N; Dickinson et al. 2003) and COSMOS fields, including photometric predictions for the MIRI F770W band. Apart from considering the three components -- stellar, AGN and star forming-IR templates -- in the fitting process, an important highlight of \textsc{Stardust} is that it treats the stellar and dust emission components independently without imposing any energy balance between the UV/optical absorption and emission in the IR.

Model predictions from \citet{Bisigello2021} are based on the SPRITZ simulation, which is built from a set of observed luminosity functions (LFs) at IR wavelengths (Herschel and \emph{K}-band) as well as a galaxy stellar mass function, covering different galaxy populations. The limitations of semi-analytical and hydrodynamical simulations in reproducing the observed number density of luminous IR galaxies is addressed in SPRITZ by properly accounting for the contribution of nebular emission arising from star formation as well as AGN to the resulting IR emission.

The \citet{Rowan-Robinson2024} count predictions are derived based on the semi-empirical model given by \citep{Rowan-Robinson2001,Rowan-Robinson2009} which assumes that the main features of the infrared galaxy population are largely captured by four different source templates (quiescent, starbursts, extreme starbursts and AGN), as detailed in \citet{Rowan-Robinson2001}.

Overall, the predictions based on SHARK and SPRITZ models tend to be in good agreement with each other as well as with our observed counts. The \citet{Cowley2018} and \citet{Rowan-Robinson2024} models appear to slightly overpredict fainter objects when compared to the other the SHARK and SPRITZ models as well as the observed counts, while the predictions from \citet{Kokorev2021} underestimates the counts below $S_\nu \sim 10\mu$Jy in comparison with all other models/observations. With increasing small number statistics towards the bright end of the observed counts, the model predictions are almost always higher in comparison. As noted in previous studies \citep[e.g.,][]{Yang2023b,Stone2024}, galaxy SEDs in the wavelength range probed by MIRI are strongly affected by PAH features which are represented schematically in the \citet{Cowley2018} and \citet{Rowan-Robinson2024} models. However, the SHARK and SPRITZ models do include an accurate representation of PAH emission and absorption features based on Spitzer IRS spectra, which might play a critical role in the resulting agreement of model predictions with the observed counts. Considering that number count estimates are based on sources across a wide range of redshifts and with different physical properties, a reasonable agreement between different model predictions and observations is promising, especially given the simplistic assumptions and differing prescriptions of galaxy properties adopted by modelling approaches. It is worth noting that such a comparison with new and powerful JWST observations, covering a wide range in flux and wavelength, can help better constrain the models of galaxy formation and evolution. 

\section{Summary and Conclusions}\label{sec:summary}
In this work, we present an overview of the observations and data reduction of MIRI F770W images from the largest Cycle 1 (GO) JWST program, COSMOS-Web. We performed our imaging reduction using the standard JWST Calibration Pipeline with an additional custom step to mitigate the characteristic MIRI background effect as detailed in Section \ref{sec:stage3}. The measured average 5$\sigma$ image depths across MIRI visits are significantly better than ETC predictions, thanks to our careful background subtraction procedure. We derive a source catalog using \texttt{SE} based on just the F770W observations and conduct simulations to measure the completeness of our sample as a function of source flux. 

We compare the JWST/MIRI F770W fluxes with archival Spitzer/IRAC CH4 photometry and find them to be largely consistent with a tiny offset that is mostly due to differing broadband transmissions. We also derive the MIRI number counts at 7.7$\mu$m and compare them to other MIRI F770W and IRAC CH4 estimates from previous studies as well as theoretical/semi-analytical model predictions from literature. Our findings show strong agreement with the estimates from various JWST programs, though there is a slight underprediction when compared to IRAC-based estimates. Notably, our number counts span an impressive range of fluxes -- five orders of magnitude, from approximately 0.2 to 2300 $\mu$Jy, which is a testament to the preeminence of COSMOS-Web.


\begin{acknowledgments}
This work is based on observations made with the NASA/ESA/CSA James Webb Space Telescope. The data were obtained from the Mikulski Archive for Space Telescopes at the Space Telescope Science Institute, which is operated by the Association of Universities for Research in Astronomy, Inc., under NASA contract NAS 5-03127 for JWST. These observations are associated with program \#1727. Support for this work was provided by NASA through grant JWST-GO-01727 awarded by the Space Telescope Science Institute, which is operated by the Association of Universities for Research in Astronomy, Inc., under NASA contract NAS 5-26555. This work was made possible by utilizing the CANDIDE cluster at the Institut d’Astrophysique de Paris, which was funded through grants from the PNCG, CNES, DIM-ACAV, and the Cosmic Dawn Center and maintained by S. Rouberol. The French contingent of the COSMOS team is partly supported by the Centre National d’Etudes Spatiales (CNES). This research made use of Photutils, an Astropy package for
detection and photometry of astronomical sources \citep{Bradley2023}.
\end{acknowledgments}

%

\facilities{JWST (MIRI)}


\software{Astropy \citep{astropy:2013, astropy:2018, astropy:2022}, ComEst \citep{Chiu2016}, matplotlib \citep{Hunter2007}, NumPy \citep{Harris2020}, photutils \citep{Bradley2023}, SciPy \citep{Virtanen2020}, SourceExtractor\citep{Bertin1996}, SourceXtractor++\citep{Bertin2020}}



\appendix
\restartappendixnumbering

\section{Summary of COSMOS-Web MIRI visits}

\begin{figure*}[]
    \centering
    \includegraphics[width=0.75\columnwidth]{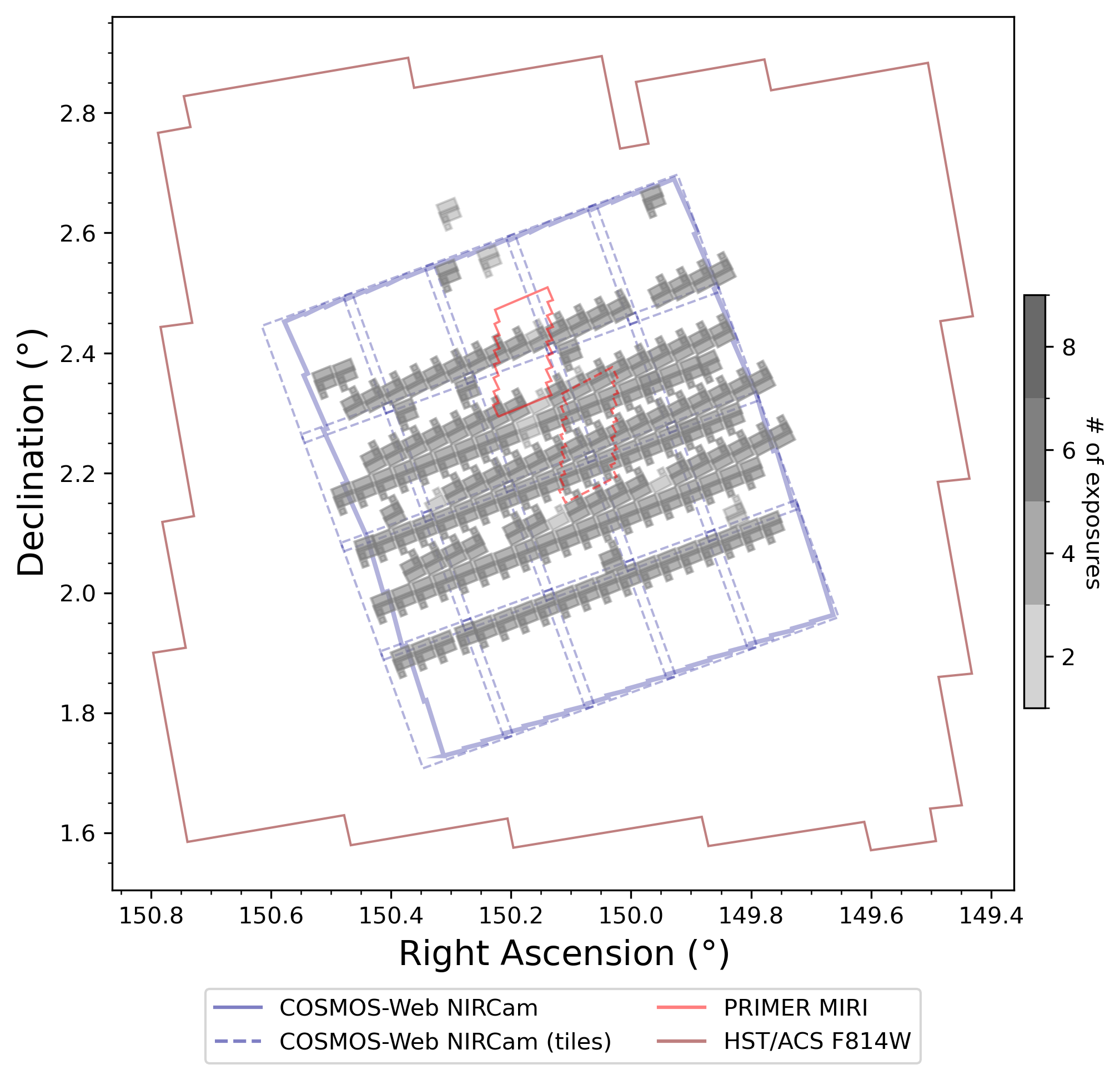}
    \caption{Exposure map of completed MIRI visits from the COSMOS-Web program. Shown for reference are (a) the HST ACS F814W mosaic outline (maroon) of the COSMOS field \citep{Koekemoer2007,Scoville2007}, (b) the overall COSMOS-Web NIRCam mosaic outline as well as the \textsf{tiles} layout (blue; Franco et al., in prep.), and (c) the PRIMER MIRI coverage (red; JWST GO PID 1837).}
    \label{fig:MIRI_visits_map}
\end{figure*}

Here we provide information on the 165 visits with MIRI F770W data that make up the COSMOS-Web MIRI coverage. Table \ref{tab:visits} lists each individual visit along with their reference positions, number of exposures, and their position angles (P.A.). These include the original visits as well as the repeats that were taken for those which either failed or were skipped due to guide star failures. Since MIRI observations were executed in parallel mode, a different P.A. for a particular repeat visit compared to the original visit meant a different region of sky coverage in MIRI. Figure \ref{fig:MIRI_visits_map} shows the exposure map of all MIRI visits listed in Table \ref{tab:visits}.\\

\twocolumngrid
\startlongtable
\begin{deluxetable}{ccccc}
  \tabletypesize{\footnotesize}
  \tablecolumns{5}
  \tablecaption{COSMOS-Web MIRI visit summary}
  \tablehead{
    \colhead{Obs No.} & \colhead{R.A.} & \colhead{Dec.} & \colhead{N$_{\textbf{exp}}$} & \colhead{P.A.} 
  }
  \startdata
        1 & 09:59:23.315 & +02:32:04.48 & 8 & 293\\ 
        3 & 09:59:31.518 & +02:31:19.75 & 8 & 293\\ 
        4 & 09:59:23.594 & +02:25:53.69 & 8 & 293\\ 
        5 & 09:59:39.710 & +02:30:35.08 & 8 & 293\\ 
        6 & 09:59:31.793 & +02:25:08.97 & 8 & 293\\ 
        7 & 09:59:47.919 & +02:29:50.30 & 8 & 293\\ 
        8 & 09:59:40.008 & +02:24:24.26 & 8 & 293\\ 
        10 & 09:59:48.203 & +02:23:39.46 & 8 & 293\\ 
        11 & 10:00:04.309 & +02:28:20.92 & 8 & 293\\ 
        12 & 09:59:56.403 & +02:22:54.72 & 8 & 293\\ 
        13 & 10:00:12.513 & +02:27:36.19 & 8 & 293\\ 
        14 & 10:00:04.592 & +02:22:10.08 & 8 & 293\\ 
        15 & 10:00:20.721 & +02:26:51.40 & 8 & 293\\ 
        16 & 10:00:12.792 & +02:21:25.37 & 8 & 293\\ 
        17 & 10:00:28.920 & +02:26:06.66 & 8 & 293\\ 
        18 & 10:00:20.776 & +02:20:32.97 & 6 & 293\\ 
        19 & 10:00:36.891 & +02:25:14.46 & 6 & 293\\ 
        20 & 10:00:29.187 & +02:19:55.93 & 8 & 293\\ 
        21 & 10:00:45.307 & +02:24:37.30 & 8 & 293\\ 
        22 & 10:00:36.738 & +02:18:48.47 & 2 & 293\\ 
        23 & 10:00:53.507 & +02:23:52.57 & 8 & 293\\ 
        24 & 10:00:45.586 & +02:18:26.47 & 8 & 293\\ 
        25 & 10:01:01.716 & +02:23:07.77 & 8 & 293\\ 
        26 & 10:00:53.792 & +02:17:41.68 & 8 & 293\\ 
        27 & 10:01:09.906 & +02:22:23.15 & 8 & 293\\ 
        28 & 10:01:01.982 & +02:16:57.04 & 8 & 293\\ 
        29 & 10:01:18.116 & +02:21:38.34 & 8 & 293\\ 
        30 & 10:01:10.182 & +02:16:12.31 & 8 & 293\\ 
        31 & 10:01:26.311 & +02:20:53.63 & 8 & 293\\ 
        32 & 10:01:18.378 & +02:15:27.60 & 8 & 293\\ 
        33 & 10:01:34.510 & +02:20:08.91 & 8 & 293\\ 
        34 & 10:01:26.586 & +02:14:42.80 & 8 & 293\\ 
        35 & 10:01:42.700 & +02:19:24.26 & 8 & 293\\ 
        36 & 10:01:34.782 & +02:13:58.06 & 8 & 293\\ 
        37 & 10:01:50.906 & +02:18:39.45 & 8 & 293\\ 
        38 & 10:01:42.981 & +02:13:13.34 & 8 & 293\\ 
        39 & 09:59:07.488 & +02:21:12.21 & 8 & 293\\ 
        40 & 09:58:59.571 & +02:15:46.13 & 8 & 293\\ 
        41 & 09:59:15.687 & +02:20:27.49 & 8 & 293\\ 
        42 & 09:59:07.761 & +02:15:01.45 & 8 & 293\\ 
        43 & 09:59:23.878 & +02:19:42.85 & 8 & 293\\ 
        44 & 09:59:15.970 & +02:14:16.65 & 8 & 293\\ 
        45 & 09:59:32.079 & +02:18:58.27 & 8 & 293\\ 
        46 & 09:59:24.171 & +02:13:31.91 & 8 & 293\\ 
        47 & 09:59:40.283 & +02:18:13.35 & 8 & 293\\ 
        48 & 09:59:32.367 & +02:12:47.21 & 8 & 293\\ 
        49 & 09:59:48.473 & +02:17:28.70 & 8 & 293\\ 
        50 & 09:59:40.557 & +02:12:02.59 & 8 & 293\\ 
        51 & 09:59:56.672 & +02:16:43.99 & 8 & 293\\ 
        52 & 09:59:48.106 & +02:10:55.16 & 2 & 293\\ 
        53 & 10:00:04.871 & +02:15:59.26 & 8 & 293\\ 
        54 & 09:59:56.961 & +02:10:33.09 & 8 & 293\\ 
        55 & 10:00:13.080 & +02:15:14.46 & 8 & 293\\ 
        56 & 10:00:05.161 & +02:09:48.34 & 8 & 293\\ 
        57 & 10:00:21.276 & +02:14:29.73 & 8 & 293\\ 
        58 & 10:00:13.350 & +02:09:03.70 & 8 & 293\\ 
        59 & 10:00:29.467 & +02:13:45.08 & 8 & 293\\ 
        60 & 10:00:21.555 & +02:08:18.91 & 8 & 293\\ 
        61 & 10:00:37.666 & +02:13:00.36 & 8 & 293\\ 
        62 & 10:00:29.097 & +02:07:11.53 & 2 & 293\\ 
        63 & 10:00:45.862 & +02:12:15.63 & 8 & 293\\ 
        64 & 10:00:37.951 & +02:06:49.47 & 8 & 293\\ 
        65 & 10:00:54.062 & +02:11:30.91 & 8 & 293\\ 
        66 & 10:00:46.138 & +02:06:04.82 & 8 & 293\\ 
        67 & 10:01:02.259 & +02:10:46.20 & 8 & 293\\ 
        69 & 10:01:10.468 & +02:10:01.41 & 8 & 293\\ 
        70 & 10:01:02.535 & +02:04:35.37 & 8 & 293\\ 
        71 & 10:01:18.013 & +02:08:53.98 & 2 & 293\\ 
        72 & 10:01:10.731 & +02:03:50.64 & 8 & 293\\ 
        74 & 10:01:18.939 & +02:03:05.85 & 8 & 293\\ 
        75 & 10:01:35.058 & +02:07:47.25 & 8 & 293\\ 
        76 & 10:01:27.134 & +02:02:21.13 & 8 & 293\\ 
        78 & 09:59:19.483 & +02:17:44.78 & 8 & 107\\ 
        79 & 09:59:35.599 & +02:22:26.18 & 8 & 107\\ 
        80 & 09:59:27.688 & +02:16:59.98 & 8 & 107\\ 
        81 & 09:59:43.793 & +02:21:41.47 & 8 & 107\\ 
        82 & 09:59:36.367 & +02:16:38.92 & 2 & 107\\ 
        83 & 09:59:52.165 & +02:21:04.51 & 6 & 107\\ 
        84 & 09:59:44.078 & +02:15:30.64 & 8 & 107\\ 
        85 & 10:00:00.165 & +02:20:12.16 & 8 & 107\\ 
        86 & 09:59:52.272 & +02:14:45.90 & 8 & 107\\ 
        87 & 10:00:08.397 & +02:19:27.21 & 8 & 107\\ 
        88 & 10:00:00.478 & +02:14:01.11 & 8 & 107\\ 
        89 & 10:00:16.589 & +02:18:42.57 & 8 & 107\\ 
        90 & 10:00:08.669 & +02:13:16.45 & 8 & 107\\ 
        91 & 10:00:24.782 & +02:17:57.84 & 8 & 107\\ 
        92 & 10:00:16.873 & +02:12:31.66 & 8 & 107\\ 
        93 & 10:00:32.981 & +02:17:13.10 & 8 & 107\\ 
        94 & 10:00:25.072 & +02:11:46.93 & 8 & 107\\ 
        95 & 10:00:41.177 & +02:16:28.38 & 4 & 107\\ 
        96 & 10:00:33.258 & +02:11:02.27 & 8 & 107\\ 
        97 & 10:00:49.385 & +02:15:43.59 & 8 & 107\\ 
        98 & 10:00:41.463 & +02:10:17.49 & 8 & 107\\ 
        99 & 10:00:57.572 & +02:14:58.96 & 8 & 107\\ 
        100 & 10:00:49.662 & +02:09:32.76 & 8 & 107\\ 
        101 & 10:01:05.777 & +02:14:14.15 & 8 & 107\\ 
        102 & 10:00:57.859 & +02:08:48.04 & 8 & 107\\ 
        103 & 10:01:13.967 & +02:13:29.51 & 8 & 107\\ 
        104 & 10:01:06.046 & +02:08:03.39 & 8 & 107\\ 
        105 & 10:01:22.166 & +02:12:44.69 & 8 & 107\\ 
        106 & 10:01:14.250 & +02:07:18.60 & 8 & 107\\ 
        107 & 10:01:30.362 & +02:11:59.95 & 8 & 107\\ 
        108 & 10:01:22.444 & +02:06:33.89 & 8 & 107\\ 
        109 & 10:01:38.567 & +02:11:15.17 & 8 & 107\\ 
        110 & 10:01:30.641 & +02:05:49.15 & 8 & 107\\ 
        111 & 10:01:46.759 & +02:10:30.41 & 8 & 107\\ 
        112 & 10:01:38.840 & +02:05:04.32 & 8 & 107\\ 
        113 & 10:01:54.956 & +02:09:45.84 & 8 & 107\\ 
        114 & 10:01:45.821 & +02:04:30.10 & 8 & 104\\ 
        115 & 09:59:11.568 & +02:12:18.67 & 8 & 107\\ 
        116 & 09:59:03.652 & +02:06:52.54 & 8 & 107\\ 
        117 & 09:59:19.764 & +02:11:33.97 & 8 & 107\\ 
        118 & 09:59:11.866 & +02:06:07.67 & 8 & 107\\ 
        119 & 09:59:27.963 & +02:10:49.24 & 8 & 107\\ 
        120 & 09:59:20.057 & +02:05:23.04 & 8 & 107\\ 
        121 & 09:59:36.159 & +02:10:04.51 & 8 & 107\\ 
        122 & 09:59:28.250 & +02:04:38.32 & 8 & 107\\ 
        123 & 09:59:44.359 & +02:09:19.77 & 8 & 107\\ 
        124 & 09:59:36.451 & +02:03:53.63 & 8 & 107\\ 
        125 & 09:59:52.554 & +02:08:35.07 & 8 & 107\\ 
        126 & 09:59:44.648 & +02:03:08.88 & 8 & 107\\ 
        127 & 10:00:00.750 & +02:07:50.36 & 8 & 107\\ 
        128 & 09:59:52.843 & +02:02:24.16 & 8 & 107\\ 
        129 & 10:00:08.949 & +02:07:05.64 & 8 & 107\\ 
        130 & 10:00:01.042 & +02:01:39.29 & 8 & 107\\ 
        131 & 10:00:17.153 & +02:06:20.84 & 8 & 107\\ 
        132 & 10:00:09.227 & +02:00:54.80 & 8 & 107\\ 
        133 & 10:00:25.349 & +02:05:36.10 & 8 & 107\\ 
        134 & 10:00:17.430 & +02:00:09.98 & 8 & 107\\ 
        135 & 10:00:33.545 & +02:04:51.37 & 8 & 107\\ 
        136 & 10:00:25.631 & +01:59:25.13 & 8 & 107\\ 
        137 & 10:00:41.731 & +02:04:06.74 & 8 & 107\\ 
        138 & 10:00:33.825 & +01:58:40.52 & 8 & 107\\ 
        139 & 10:00:49.940 & +02:03:21.93 & 8 & 107\\ 
        140 & 10:00:42.021 & +01:57:55.79 & 8 & 107\\ 
        141 & 10:00:58.126 & +02:02:37.31 & 8 & 107\\ 
        142 & 10:00:50.209 & +01:57:11.17 & 8 & 107\\ 
        143 & 10:01:06.321 & +02:01:52.56 & 8 & 107\\ 
        144 & 10:00:58.401 & +01:56:26.38 & 8 & 107\\ 
        145 & 10:01:14.518 & +02:01:07.84 & 8 & 107\\ 
        146 & 10:01:05.656 & +01:55:50.11 & 8 & 105\\ 
        147 & 10:01:22.713 & +02:00:23.13 & 8 & 107\\ 
        148 & 10:01:14.792 & +01:54:57.01 & 8 & 107\\ 
        149 & 10:01:30.917 & +01:59:38.26 & 8 & 107\\ 
        150 & 10:01:22.988 & +01:54:12.29 & 8 & 107\\ 
        151 & 10:01:39.113 & +01:58:53.59 & 8 & 107\\ 
        152 & 10:01:31.183 & +01:53:27.57 & 8 & 107\\ 
        \hline
        \multicolumn{5}{c}{Repeat visits}\\
        \hline
        153 & 09:59:28.769 & +02:22:57.38 & 8 & 110\\ 
        154 & 09:59:36.367 & +02:16:38.92 & 4 & 107\\  
        155 & 09:59:18.022 & +02:07:46.70 & 4 & 289\\
        156 & 10:00:07.665 & +02:03:13.85 & 8 & 288\\
        157 & 09:59:52.165 & +02:21:04.51 & 4 & 107\\
        158 & 09:59:51.153 & +02:39:29.11 & 8 & 107\\
        159 & 10:00:56.748 & +02:33:31.31 & 4 & 107\\ 
        160 & 10:01:12.869 & +02:38:12.71 & 4 & 107\\ 
        161 & 10:01:13.147 & +02:32:01.88 & 8 & 107\\ 
        162 & 10:00:24.513 & +02:24:08.52 & 8 & 107\\ 
        163 & 10:01:05.501 & +02:20:24.92 & 8 & 107\\ 
        164 & 10:01:30.092 & +02:18:10.77 & 8 & 107\\ 
        165 & 10:01:54.410 & +02:22:07.43 & 8 & 107\\ 
        166 & 10:02:02.605 & +02:21:22.72 & 8 & 107\\
        167 & 09:59:36.367 & +02:16:38.92 & 4 & 107\\ 
  \enddata
  \label{tab:visits}
  \tablecomments{Visits \#2, \#68, \#73 and \#77 were fully skipped, hence no corresponding entries in the above table. Repeat visits \#154 and \#167 have same reference positions as visit \#82 and, repeat visit \#167 has same reference position as visit \#83 since these were overlapping with the corresponding original visit.
  }
\end{deluxetable}

\bibliography{refs}{}
\bibliographystyle{aasjournal}



\end{document}